\documentclass[%
 reprint,
 amsmath,amssymb,
prl,
]{revtex4-2}

\usepackage{lipsum}
\usepackage{graphicx}
\usepackage{bm}
\usepackage{color}
\usepackage{times}
\usepackage{amsmath}
\usepackage{float}
\usepackage{hyperref}
\usepackage{ulem} 
\usepackage[dvipsnames]{xcolor}

\newcommand{\beginendmatter}{%
	\setcounter{table}{0}
	\renewcommand{\thetable}{A\arabic{table}}%
	\setcounter{figure}{0}
	\renewcommand{\thefigure}{A\arabic{figure}}%
 	\setcounter{equation}{0}
	\renewcommand{\theequation}{A\arabic{equation}}%
	\setcounter{page}{0}
	\renewcommand{\thepage}{A\arabic{page}}%
	
}

\newcommand{\beginsupplement}{%
	\setcounter{table}{0}
	\renewcommand{\thetable}{S\arabic{table}}%
	\setcounter{figure}{0}
	\renewcommand{\thefigure}{S\arabic{figure}}%
 	\setcounter{equation}{0}
	\renewcommand{\theequation}{S\arabic{equation}}%
	\setcounter{page}{1}
	\renewcommand{\thepage}{S\arabic{page}}%
	
}

\begin{document}


\title{Topological mode conservation and conversion in phononic crystals with temporal interfaces}

\author{Mahmoud M. Samak}
\author{Osama R. Bilal}
\email{osama.bilal@uconn.edu}
 \affiliation{School of Mechanical, Aerospace, and Manufacturing Engineering, University of Connecticut, Storrs , CT , 06269, USA.}

\date{\today}

\begin{abstract}
A sudden change in material properties creates a temporal interface and forces a propagating wave to change its frequency while preserving its wavenumber. In contrast to monoatomic lattices with a single frequency-wavenumber pair, polyatomic lattices support multiple frequencies for each wavenumber. To date, experimental observations are limited to topologically trivial monoatomic phononic systems. Here, we utilize analytical, numerical, and experimental methods to examine topologically non-trivial phononic lattices subject to temporal interfaces. In particular, we realize phononic lattices demonstrating single-frequency shift (i.e., mode conservation) and multi-frequency splitting (i.e., mode conversion) following a temporal interface. Accordingly, we generalize temporal analogues of Snell's law and Fresnel equations. Moreover, we utilize Bloch mode overlaps to obtain a phononic time lens and a classical analogue of dynamic quantum phase transitions for phonons. Such overlap determines the probability of mode conversion or conservation after a temporal interface and, more importantly, can carry hidden topological characteristics. Our methodology paves the way for the use of temporal interfaces in probing phonon band topology and the realization of advanced acoustic devices.
\end{abstract}

\maketitle



Time-varying media utilize time as an additional dimension by modulating material properties periodically \cite{sounas2017non, wang2018observation,wallen2019nonreciprocal, nassar2020nonreciprocity, lee2021parametric,  kitagawa2010topological, wang2020floquet} or aperiodically \cite{caloz2019spacetime, caloz2019spacetime2} in time. An intriguing class of time-varying media exhibits a sudden change in material properties over a relatively short period of time (i.e., temporal interface) \cite{morgenthaler1958velocity, xiao2014reflection, lee2018linear, zhou2020broadband,long2023time, moussa2023observation}. Due to the breaking of time translational symmetry (TTS), a propagating wave experiences time-reflection and time-refraction at a temporal interface accompanied by a change in its frequency, but a conservation of its wavenumber \cite{mendoncca2000quantum, mendoncca2002time, mendoncca2005time, biancalana2007dynamics, plansinis2015temporal, gratus2021temporal, zhang2021temporal}. Such behavior is the temporal analogue of wave scattering at spatial interfaces, supporting the theory of space-time duality \cite{akhmanov1969nonstationary,kolner2002space}. Experimental observations of frequency conversion at temporal interfaces are universal across different wave types, spanning electromagnetic waves \cite{wilks1988frequency, nishida2012experimental, moussa2023observation, jones2024time}, water waves \cite{bacot2016time,apffel2022frequency}, optical waves \cite{lee2018linear,zhou2020broadband,bohn2021spatiotemporal,miyamaru2021ultrafast}, acoustic waves \cite{kim2024temporal}, elastic waves \cite{liu2024inherent,wang2025temporal}, and quantum systems \cite{dong2024quantum}. These advances in understanding temporal interfaces led to the realization of anti-reflection temporal coatings \cite{pacheco2020antireflection}, thin absorbers \cite{li2020temporal}, instantaneous time-mirrors \cite{bacot2016time,fink2017time}, temporal super-lensing \cite{ye2023reconfigurable}, wave freezing and reversing \cite{karki2021stopping,riva2025non}, broadband temporal coherent wave control \cite{galiffi2023broadband}, and temporal reflection control \cite{zhang2021impact,santini2023elastic}. In addition, different aspects of temporal symmetry have been considered, including temporal-spatial interfaces \cite{delory2024elastic,pacheco2025temporal}, temporal layered media \cite{torrent2018loss, lu2018time, trainiti2019time, ramaccia2020light, castaldi2021exploiting, ramaccia2021temporal, li2022nonreciprocity, dong2024temporal}, the sudden activation or deactivation of spatio-temporal modulation \cite{ye2025nonreciprocal}, and the topological characteristics associated with temporal interface \cite{wu2024edge,xu2025probing}. Although recent advances in photonics have shown that temporal interfaces can probe topological phase transition \cite{wu2024edge,xu2025probing}, the literature on phononic lattices remains limited to topologically trivial (i.e., monoatomic) lattices \cite{kim2024temporal, liu2024inherent, wang2025temporal}. Non-trivial topology requires more lattice sites per unit cell (i.e., polyatomic) to observe a temporal analogue of spatial topological insulators. However, for polyatomic system, a wave packet has more than one possible dispersion branch to ``hop" to while preserving its wavenumber, begging the question, what is the probability distribution for each mode following a temporal interface. 
In this letter, we utilize analytical, numerical and experimental methods to investigate spatially-homogeneous polyatomic lattices subject to temporal interfaces. We establish a design methodology to engineer Bloch-mode overlaps to determine the probability of mode conversion or conservation after a temporal interface. Moreover, we harness the overlaps to demonstrate both a phononic time lens and a classical analogue of dynamic quantum phase transitions for phonons.  

We start our analysis by considering a \textit{polyatomic} lattice subjected to a temporal interface between two homogeneous states ($S_1$) and ($S_2$). Each unit cell consists of $n$ masses, $n$ inter-stiffness, and $n$ ground springs. Hence, each state has $n$ dispersion branches. To quantify the probability of conversion ($|\Psi_{i,j}|^2$) of a mode $i$ in state $S_1$ to a mode $j$ in state $S_2$, we calculate the Bloch mode shape overlap, $\Psi_{i,j}=\mathbf{\phi_{S1}^i}(\kappa_0)^\dag \mathbf{M} \mathbf{\phi_{S2}^j} (\kappa_0)$, where $\mathbf{M}$ is the mass matrix, \textbf{ $\phi_{S1}^i$} is mode $i$ at state $S_1$ before the interface and \textbf{$\phi_{S2}^j$} is mode $j$ at state $S_2$ after the interface, $\kappa_0$ is the wavenumber corresponding to the excitation frequency, and  $(.)^\dag$ is the complex conjugate transpose. The sum of the conversion probability for a given mode $i$ in state $S_1$ is $\sum_{j=1}^n|\Psi_{i,j}|^2=1$ ($j=1,2,...,n$). In the case of uniform breaking of TTS, a single mode $l$ from state $S_2$ matches the originally excited mode $i$ from $S_1$ leading to mode conservation, $\phi_{S1}^i$=$\phi_{S2}^l$,   $\Psi_{i,l}=1$ and $\Psi_{i,j\ne l}=0$.  In the case of non-uniform TTS breakage by the temporal interface, the excited mode at state ($S_1$) is split between multiple modes in state $S_2$. Within our framework, we formulate the necessary conditions to satisfy uniform breaking of TTS [See Appendix A and Figs.\ref{fig:methodology}-\ref{fig:methodology_4}].

 \begin{figure} [!t]
\includegraphics [width = \columnwidth]{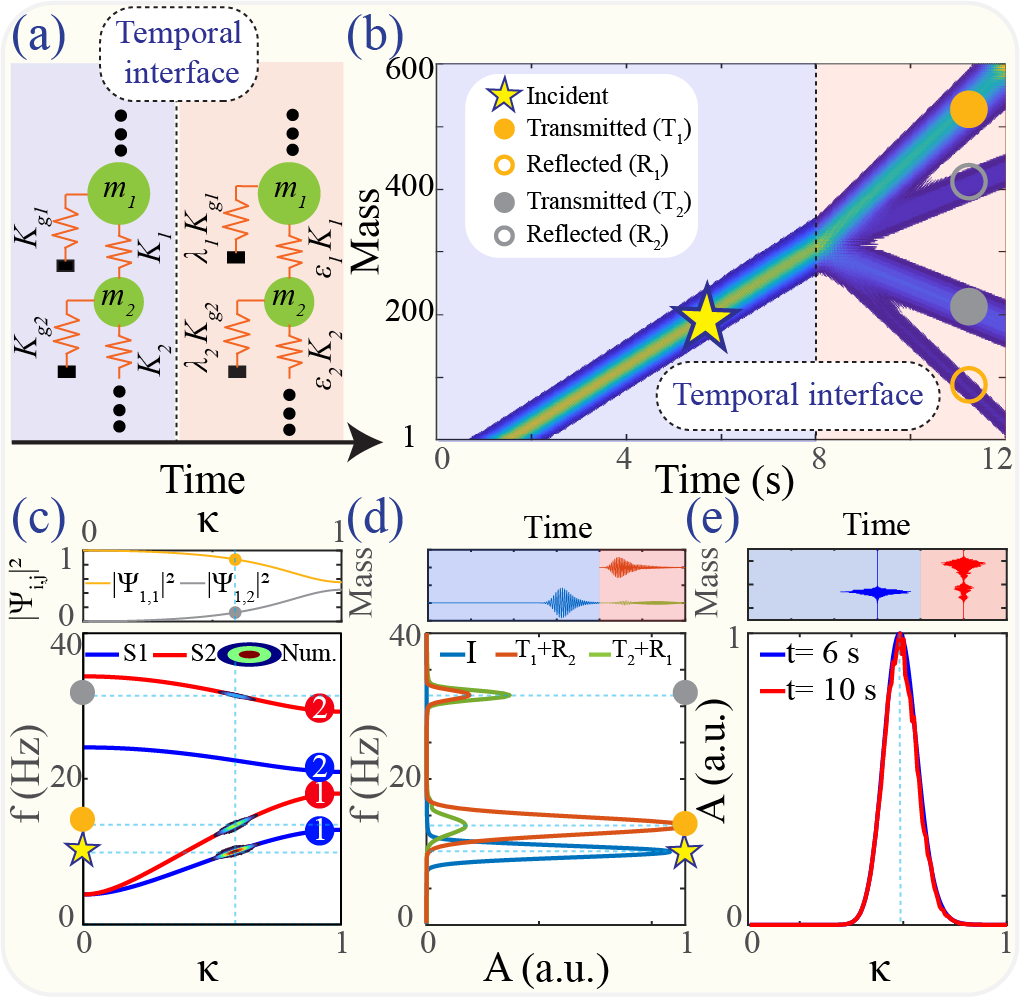} 
\caption{\label{fig:concept}\textbf{Mode conversion after temporal interfaces.} (a) A temporal interface between two diatomic lattices ( $m_1$=0.5~$m_2$=0.1Kg, $K_1$=0.5~$K_2$=500 N/m, $K_{g1}$=$K_{g2}$=100 N/m). (b) A wave packet with incident frequency 10 Hz propagates in $S_1$. At t=8 s, a temporal boundary changes the structure from $S_1$ to $S_2$. Two pairs of wave packets appear after TI (T1 and R2 to forward while T2 and R1 to backward). (c) (top) Analytical probability of each mode in $S_2$ corresponding to an incident wave on mode 1 of $S_1$. (bottom) Dispersion curves of each state with numerical 2D FFT overlay.  (d) (top) Time response of (mass 200 before TI), (mass 400 after TI) and (mass 200 after TI) showing incident, forward, and backward waves, respectively. (bottom) FFT of the three time signals in the top panel showing frequency conversion to multiple frequencies.  (e) (top) Spatial profiles at t=6 s (before TI) and t=10 s (after TI). (bottom) FFT of the two spatial profiles in the top panel showing conservation of wavenumber.  
}
\end{figure}
 
To demonstrate our theory, we simulate a diatomic structure with two masses $m_1$ and $m_2$ per unit cell. Initially, the masses are connected to each other through inter-stiffness $K_1$ and $K_2$ and to the ground through $K_{g1}$ and $K_{g2}$ in state $S_1$. We excite a homogeneous finite structure with $N$ masses in state $S_1$, with a wave packet centered at $10$ Hz. Once the wave packet reaches the middle of the finite structure, at time $t=\tau$, we introduce a temporal interface, switching the system to state $S_2$, by changing the inter-stiffness to $\epsilon_1~K_1$ and $\epsilon_2 K_2$ and  the ground stiffness to $\lambda_1~K_{g1}$ and $\lambda_2~K_{g2}$  (Fig.\ref{fig:concept}-a). To examine the general case, we design the interface to break TTS non-uniformly with $\epsilon_1=4, \epsilon_2=\lambda_1=\lambda_2=1$. As a result, the original wave packet breaks into two pairs of propagating waves: two forward wave packets and two backward ones (Fig.\ref{fig:concept}-b). To better understand this behavior, we calculate the overlap probability between modes as a function of the wavenumber. At the incident wavenumber of $\kappa_0 = 0.6$, $|\Psi_{1,1}|^2=0.875$ and $|\Psi_{1,2}|^2=0.125$, which translates to a significant conversion to both modes after passing the temporal interface (Fig.\ref{fig:concept}-c). 

To further validate our results, we overlay the spatio-temporal 2D-FFT contours of the time signal after and before the interface on the analytical dispersion curves for both states: $S_1$ (blue) and $S_2$ (red) (Fig.\ref{fig:concept}-c). The numerically-obtained contours show a signature of both mode 1 and mode 2 in state $S_2$. To further visualize the results, we consider three time signals at masses: N/2-G  at $t<\tau$, N/2+G at $t>\tau$, and N/2-G at $t>\tau$, where G is an arbitrary integer, to examine incident, forward, and backward waves, respectively. We calculate the FFT of the three time signals showing the frequency conversion after the interface (Fig.\ref{fig:concept}-d). The single incident frequency of 10 Hz splits into two different frequencies: 14.3 Hz and 31 Hz. In addition, to confirm the preservation of the wavenumber, we consider the spatial profiles at two time instances before and after the temporal interface $\tau-T$ and $\tau+T$, where T is an arbitrary constant. The calculated  FFT of the spatial profiles show wavenumber conservation even when the frequency is converted to multiple different frequencies (Fig.\ref{fig:concept}-e). However, the frequency splitting suggests that we need an updated framework for both Snell's law and Fresnel equations to account for the multiple reflections and refractions at the temporal interface for polyatomic lattices [See Appendix B]. 

\begin{figure} [b]
\includegraphics [width = \columnwidth]{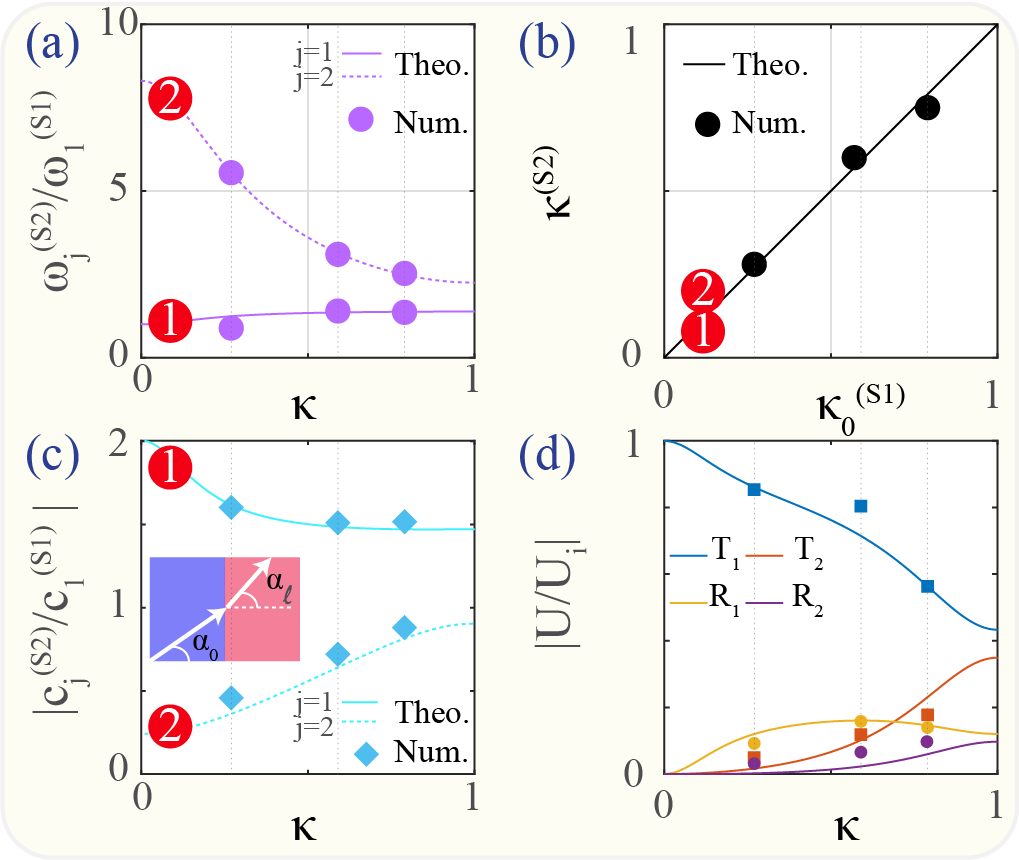} 
\caption{\label{fig:theory}\textbf{ Modification of temporal analogue of Snell's law and Fresnel equations in phononic lattices.} For the structure illustrated in Fig.\ref{fig:concept}, if the incident wave is on branch 1 of $S_1$: (a) Frequency conversion corresponding to each wavenumber and numerically calculated frequencies after TI for three different wavenumbers. (b) Numerically calculated wavenumbers before $\kappa^{(S1)}_0$ and after $\kappa^{(S2)}$ TI to show conservation of wavenumber. (c) Group velocity ratio after $C^{(S2)}_j$ and before $C^{(S1)}_1$ TI theoretically (lines) compared to the ratio between transmitted waves to incident wave angles (diamonds) to show Snell's law. (d) Ratio between transmitted ( T1 and T2) or reflected ( R1 and R2) wave packet amplitude to the incident wave packet amplitude theoretically (lines) and numerically (markers) to show Fresnel equations.}
\end{figure}
To demonstrate the generality of our framework, we test different wavenumber-frequency excitations at a temporal interface separating two homogeneous polyatomic lattices (Fig.\ref{fig:theory}).  Our results verify the conversion of frequency (Fig.\ref{fig:theory}-a) and conservation of wavenumber (Fig.\ref{fig:theory}-b) both analytically and numerically. Our updated Snell's law establishes a theoretical relation between the ratio of group velocity $C_j^{(S2)}/{C_i^{(S1)}}$ and the frequency ratio $\omega_i^{(S1)}/{\omega_j^{(S2)}}$ with a proportional constant that depends on the parameters of the two states $S_1$ and $S_2$, and the incident wavenumber. The same group velocity ratio can be obtained numerically by measuring the ratio between the transmitted wave angle and the incident wave angle. We obtain a good matching between theoretical and numerical calculations of the modified Snell's law (Fig.\ref{fig:theory}-c). To update the Fresnel equations, we use the continuity in the displacement and velocity fields before and after the temporal interface to formulate an analytical relation between each wave packet amplitude. The same ratios can be measured numerically for validation (Fig.\ref{fig:Example_Fernsel equations}).   We obtain a good match between the analytical predicted and numerically simulated models confirming our framework (Figs.\ref{fig:theory}-d and \ref{fig:Example_Fernsel equations}).   

\begin{figure}[b] 
\includegraphics [width = \columnwidth]{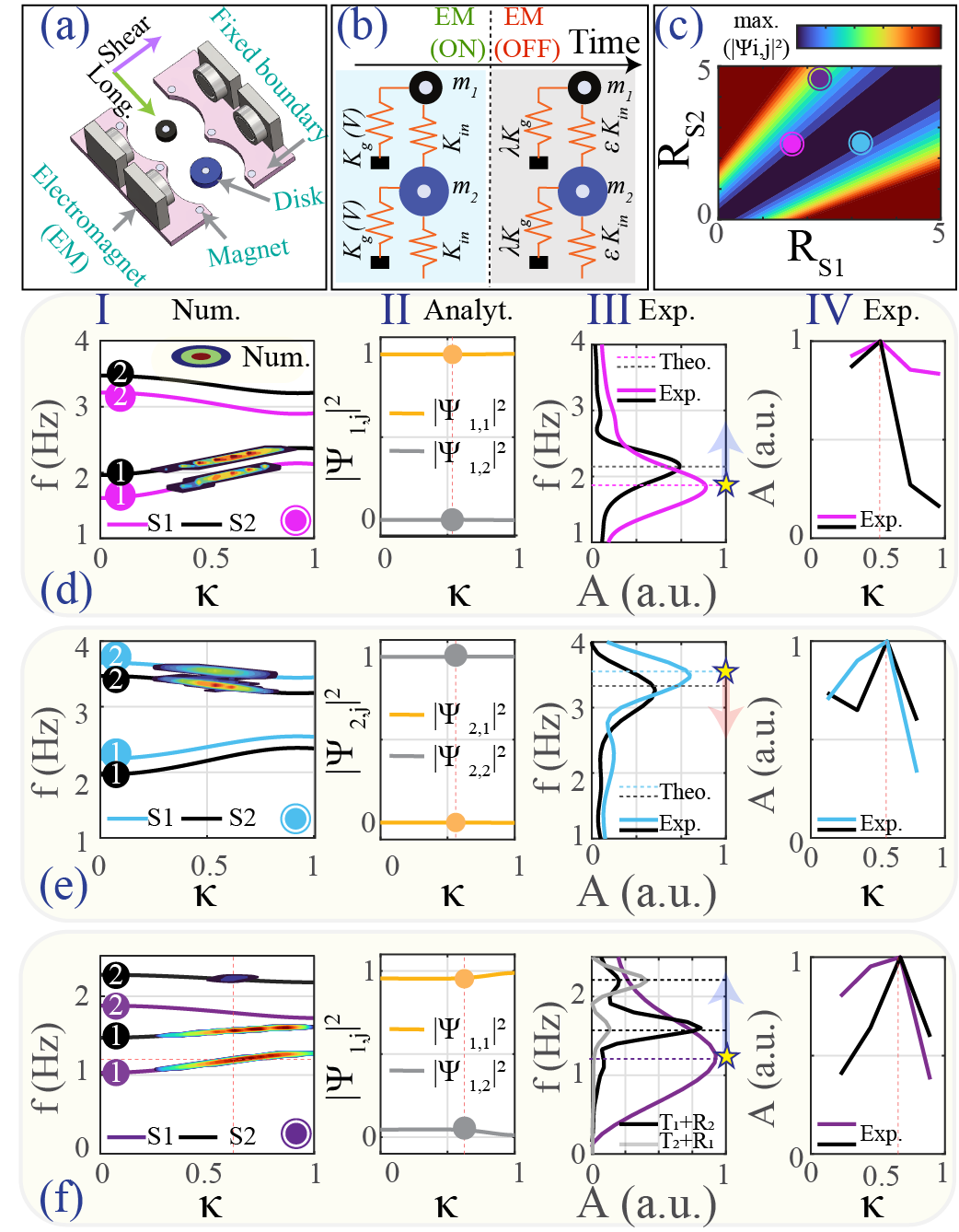} 
\caption{\label{fig:Diatomic}\textbf{Experimental observation of mode conversion.} 
(a) Physical diatomic unit cell. (b) Mathematical model. (c) Maximum obtainable overlap between acoustic and optical modes by changing ground stiffness. We show three cases: (d) changing voltage from -11 V to 0 V while a longitudinal wave is propagating, (e) changing voltage from +7 V to 0 V while a longitudinal wave is propagating, and (f) changing voltage from +5 V to 0 V while a shear wave is propagating. For each case we show (I) analytical dispersion curves and numerically calculated 2D FFT overlay, (II) analytical probability of each mode in state 2, and experimentally observed (III) frequencies and (IV) wavenumbers before and after the temporal interface.}
\end{figure}

To validate our theory experimentally, we utilize programmable polyatomic magnetic lattices composed of free floating magnetic disks. Each disk is surrounded by four fixed permanent magnets. All the permanent magnets in the boundary and disks are pointing upwards, working as repelling monopoles in-plane. The disk has two decoupled modes, longitudinal and shear. The repulsion force between the disk magnet and the boundary magnets represents the ground stiffness $K_{g-b}$ while the repulsion between the disks in each unit cell represents the inter-stiffness $K_{in}$ \cite{watkins2020demultiplexing, watkins2021exploiting, norouzi2021classification, eichelberg2022metamaterials, watkins2022harnessing, samak2024evidence,   samak2024direct, stenseng2025bi}. To change the ground stiffness in real-time, we place an electromagnet on each side of the unit cell, adding an extra term to the lattice's total ground stiffness $K_{g-EM}(V)$. (Fig.\ref{fig:Diatomic}-a and Fig.\ref{fig:setup}).  The lattice can be modeled as a spring-mass system ($m_1=0.34$ g, $m_2=0.62$ g and $K_{in,long.}=0.0306$ N/m or $K_{in,shear}=-0.0095 $ N/m) with a total ground stiffness $K_g(V)$=$K_{g-b}$+$K_{g-EM}(V)$ (Fig.\ref{fig:Diatomic}-b) [See Appendix C for more details]. 

We start our experimental validation by calculating the maximum conversion probability $|\Psi_{i,j\ne i}|^2$ as a function of the stiffness ratio ($R={K_g}/{K_{in}}$) to explore the possibility of mode conservation and conversion with our lattices (Fig.\ref{fig:Diatomic}-c). We consider three cases from the stiffness ratio map corresponding to (1) acoustic mode conservation with frequency-up conversion, (2) optical mode conservation with frequency-down conversion, and (3) optical-mode splitting into both acoustic and optical modes. For case 1, we introduce the temporal interface by changing the stiffness ratio from $R_{S1}=1.66$ to $R_{S2}=2.5$, which is equal to $\epsilon_1=\epsilon_2=1$ and $\lambda_1= \lambda_2=1.5$ (i.e., non-uniform breaking of TTS). We excite the lattice with a wave packet longitudinally at $f=1.87$ Hz and $\kappa_0=0.514\frac{\pi}{a}$  (Fig.\ref{fig:Diatomic}-d). The conversion probability between the two acoustic modes $\phi^1_{S1}$ and $\phi^1_{S2}$ is almost unity $|\Psi_{1,1}|^2\approx 1$, leading to mode conservation following the temporal interface (Fig. \ref{fig:Diatomic} d (II)). This is further verified by superimposing the numerical spatiotemporal 2D-FFT contours on the dispersion curves (Fig.\ref{fig:Diatomic} d (I)). In addition, the experimentally measured frequency (Fig.\ref{fig:Diatomic}-d(III)) and wavenumber (Fig.\ref{fig:Diatomic}-d(IV)) before and after the temporal interface show clear mode preservation with frequency-up conversion from $f^{(S1)}_1=1.87$ Hz to $f^{(S2)}_1=2.2$ Hz confirming both our analytical and numerical predictions. For case 2, the temporal interface changes the stiffness ratio from $R_{S1}=3.2$ to $R_{S2}=2.5$, which is equal to $\epsilon_1=\epsilon_2=1$ and $\lambda_1=\lambda_2=0.78$ (i.e., non-uniform breaking of TTS). We excite the lattice with a wave packet longitudinally at $f=3.55$ Hz (Fig.\ref{fig:Diatomic}-e). By design, despite the existence of two possible modes, the conversion probability between the two optical modes $\phi^2_{S1}$ and $\phi^2_{S2}$ is almost unity $|\Psi_{2,2}|^2\approx 1$, leading to optical mode conservation following the temporal interface (Fig. \ref{fig:Diatomic} e (II)). The numerical contours verify our design approach numerically and the measured frequency (Fig.\ref{fig:Diatomic} e(III)) and wavenumber (Fig.\ref{fig:Diatomic} e(IV)) provide further evidence of mode preservation with frequency-down conversion.


\begin{figure}[!t]
\includegraphics [width = \columnwidth]{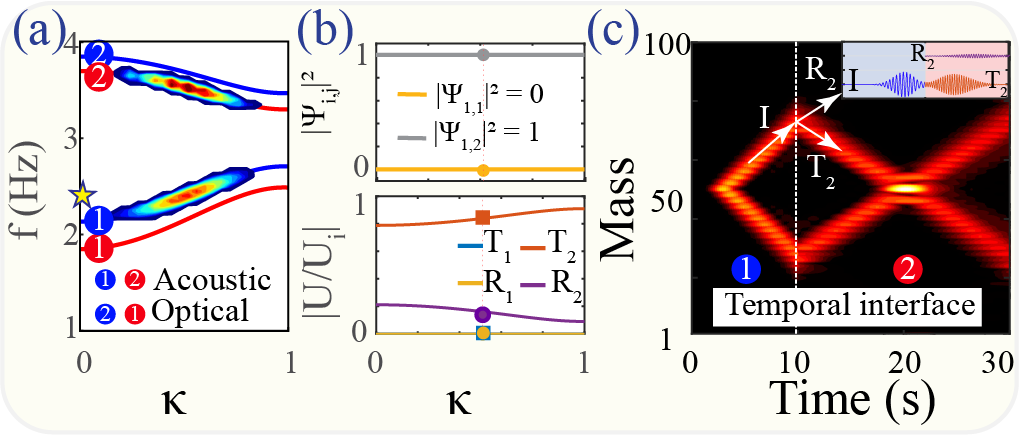} 
\caption{\label{fig:focusing}\textbf{Phononic time-lens.} A temporal interface between two diatomic lattices ( $m_1$=0.5~$m_2$=0.34 g, $K_1$=0.5~$K_2$=0.0306 N/m, $K_{g1}$=0.5$K_{g2}$=2$K_1$, $\epsilon_1=\epsilon_2=-1$ and $\lambda_1=\lambda_2=3$). (a) Analytical dispersion curves and numerically calculated 2D FFT overlay. (b) (top) Analytical probability of each mode in state 2. (bottom) Ratio between wave packets' amplitudes (c) The mid-mass is excited with a wave packet with a temporal interface after 10 s.}
\end{figure}

To demonstrate mode splitting, we introduce the time interface by changing the shear stiffness ratio from $R_{S1}=-6.2$  to $R_{S2}=-8.7$, which is equal to $\epsilon_1$= $\epsilon_2$=1 and $\lambda_1$= $\lambda_2$=1.4 (i.e., non-uniform breaking of TTS). We excite the lattice with a wave packet in the shear direction centered at $f=1.2$ Hz (Fig.\ref{fig:Diatomic}-e). This intriguing case, show probability of conversion to both modes after the temporal interface ( Fig.\ref{fig:Diatomic}-f(II)). The signature of mode splitting is evident by the superimposed 2D-FFT contours on top of the dispersion branches. It worth noting that the lower branch has higher conversion probability and therefore has a higher signature (Fig.\ref{fig:Diatomic}-f(I)). We verify the mode splitting experimentally by measuring  both the frequency (Fig.\ref{fig:Diatomic}-f(III)) and wavenumber (Fig.\ref{fig:Diatomic}-f(IV)) before and after the temporal interface. Analytical, numerical, and experimental results show very good agreement, proving  optical mode conversion following the temporal interface with uneven splitting between acoustic and optical modes based on the calculated probability. 
  
Both uniform and non-uniform breaking of time translational symmetry can have intriguing implications for propagating phonons. One of these implications could be full mode conservation despite an opposite group velocity direction with uniform symmetry breaking, leading to phonon time-lensing. Another implication is the ability to harness temporal interfaces as means of probing the bulk band topology of phonons. To demonstrate temporal lensing, we consider a diatomic lattice with a uniform breaking of TTS through a temporal interface with $\epsilon_1=\epsilon_2=-1$ and $\lambda_1=\lambda_2=3$. We note the use of negative inter-stiffness after the temporal interface, to swap between the acoustic and optical branches of our phononic crystal \cite{samak2025observation}(Fig.\ref{fig:focusing}-a). Based on our model, the frequency conversion due to the uniform breaking of TTS satisfies $(f^{(S2)})^2-\epsilon~(f^{(S1)})^2=(\lambda-\epsilon)~\frac{K_{gl}}{4~\pi ^2~m_l}$. For this case, the probability of mode 1 in state $S_1$ transforming into mode 2 in state $S_2$ is 100\%. We excite a finite structure of 100 masses with a wave packet centered at $f^{S1}_1$=2.4 Hz at mass number 50. Once the temporal interface is activated at $t = 10$s, the frequency shifts-up to $f^{S2}_2$=3.6 Hz. It is important to note that the wave hops to mode 2 despite not being the closest in frequency, nor the closest in group velocity value or sign (Fig.\ref{fig:focusing}-b). As a result of the change in group velocity sign, the waves at the temporal interface reflect back to combine with amplification (Fig.\ref{fig:focusing}-c) creating an example of a ``time lens" with uniform breaking of temporal symmetry.   

\begin{figure}[b] 
\includegraphics [width = \columnwidth]{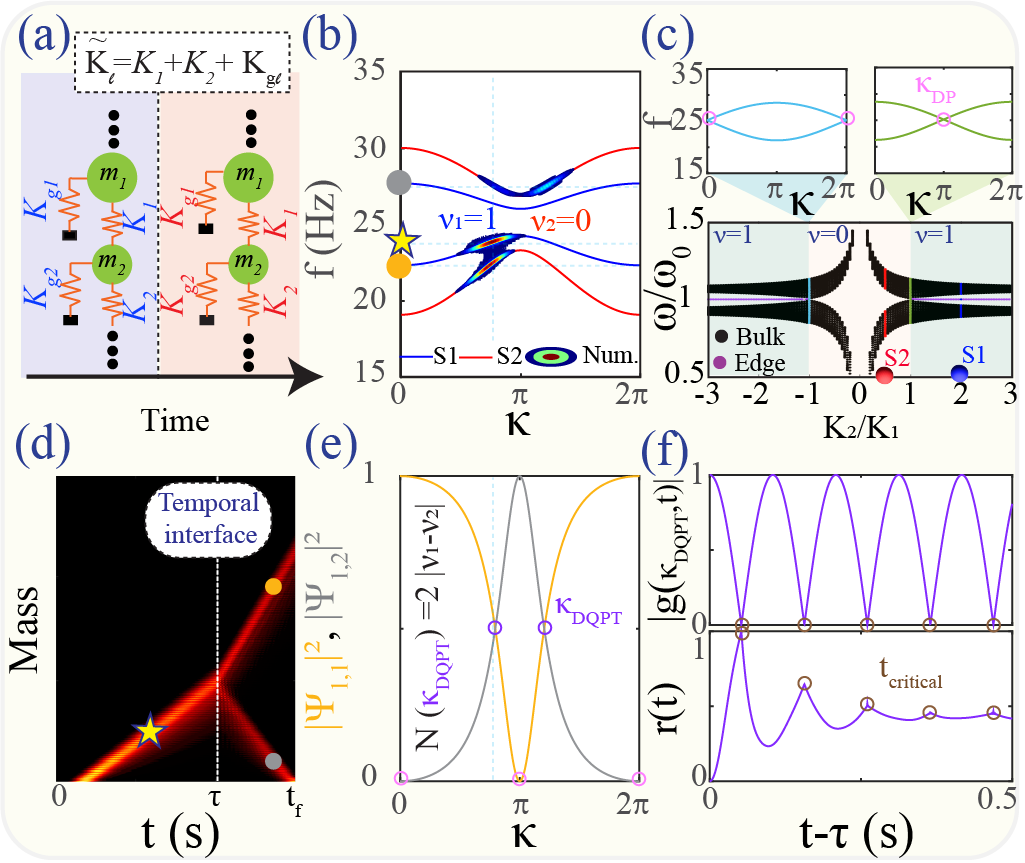} 
\caption{\label{fig:topology}\textbf{Probing bulk band topology with temporal interfaces.} (a) Mathematical model (b) Analytical dispersion curves for each state and numerical 2D FFT overlay. (c) Phase diagram shows the relation between inter-stiffnesses ratio $\frac{K_2}{K_1}$ and number of topologically protected edge modes (i.e., winding number $\nu$) with the chiral symmetry necessary condition  $m_1\Tilde{K_2}=m_2\Tilde{K_1}$. Blue (red) dot represents $S_1$ ($S_2$). The vertical lines at $K_2/K_1=1$ and $-1$ represent topological phase boundaries and the corresponding dispersion curves are shown above. (d)  The first mass is excited with a wave packet centered at $f=24$ Hz corresponding to $\kappa_{DQPT}=0.8 \frac{\pi}{a}$. A temporal interface at $t=\tau$ changes the system abruptly to $S_2$.  (e) Analytical probability of each mode in $S_2$ corresponding to an incident wave on mode 1
of $S_1$. (f)  Revival  ``Loschmidt” amplitude (top) and rate function (bottom). The revival amplitude vanishes at the critical times (circles) which are the same times corresponding to nonanalytic cusps in the rate function which represent dynamical quantum phase transition in the time boundary effect.}
\end{figure}

Non-uniform breaking of translational time symmetry can be utilized to probe bulk band topology. As a demonstration, we consider a diatomic structure in its general form. The band gap between the two dispersion branches closes when ${K_2}/{K_1}=1~ $or $-1$ only if $m_1\Tilde{K_2}=m_2\Tilde{K_1}$ which is  a  necessary condition to preserve chiral symmetry and capture topological phase boundaries ($\Tilde{K_l}=K_1+K_2+K_{gl}$). It is important to note that breaking mass symmetry in diatomic lattices prevents band gap closing even as the winding number changes from 1 to 0 or vice-versa, which breaks bulk-edge correspondence \cite{chen2018study}. In our model, we use ground stiffness as means to restore chiral symmetry, preserving the principle of bulk-edge correspondence even with different masses [See Appendix D]. The initial state for the lattice $S_1$ has ${K_2}/{K_1}=2$ and a winding number $\nu=1$. The temporal interface changes the system abruptly in a nonuniform manner to $S_2$ with ${K_2}/{K_1}=0.5$ and a winding number $\nu=0$. Both states satisfy the chiral symmetry necessary condition. As a result of the nonuniform breaking of TTS, a single incident wave splits into two frequencies at the temporal interface (Fig.\ref{fig:topology}-b,d). Topological characteristics can be probed from Bloch mode shapes overlaps (Fig.\ref{fig:topology}-e). As a result of switching between two states with different $K_2/K_1$ ratios, Bloch mode overlaps vanish at the wavenumbers of band gap closing  ($\kappa_{DP}$ or Degenerate Points) (Fig.\ref{fig:topology} c,e). Moreover, as a result of switching between topologically distinct states ($\nu_1 \ne \nu_2$), Bloch mode overlaps intersect at the dynamic quantum phase transition points ($\kappa_{DQPT}$) (Fig.\ref{fig:topology}-e). Such points ensure the existence of vanishing revival amplitude ($g(\kappa,t)$) at critical times ($t_{critical}$) which
are the same times corresponding to nonanalytic cusps in rate function ($r(t)$) (Fig.\ref{fig:topology}-f). The non-analytical behavior at these critical times is the phononic analogue of dynamical quantum phase
transitions with temporal boundaries \cite{heyl2018dynamical}.

In conclusion, a temporal interface is a sudden change in the hosting media leading to a break of temporal translational symmetry (TTS). A uniform break of TTS guarantees mode conservation and a single frequency emerges after the temporal interface. In contrast, a non-uniform breaking of TTS leads to mode conversion and the nucleation of multiple frequencies at the temporal interface. Through a combination of experimental, numerical and analytical methods, we show a design methodology for both mode conservation and mode conversion. We harness our design methodology to obtain a phononic time-lens by mode conservation and an analogue to dynamic quantum phase transition in time by mode conversion. Our work paves the way for utilizing temporal interfaces in different wave domains in a fully programmable way.      

This work was supported under Cooperative Agreement W56HZV-21-2-0001 with the US Army DEVCOM Ground Vehicle Systems Center (GVSC), through the Virtual Prototyping of Autonomy Enabled Ground Systems (VIPR-GS) program and the Air force Research Labs, Materials and Manufacturing Directorate (AFRL/RXMS) contract No. FA8650–21–C5711. DISTRIBUTION STATEMENT A. Approved for public release; distribution is unlimited.

\bibliography{references}

\newpage
\newpage
\newpage

\beginendmatter ~
\appendix
\newpage
\newpage
\section{Appendix A: Design methodology} 
A temporal interface (TI) breaks temporal translational symmetry (TTS) forcing a propagating wave to change its frequency while preserving its wavenumber. Uniform breaking of TTS translate to two states $S_1$ and $S_2$ with a shift in the frequency while conserving mode shapes at all wavenumbers (i.e., $\phi_{S1}^{l}(\kappa)=\phi^{l}_{S2}(\kappa)$). To achieve a uniform breaking of TTS and obtain a single frequency (i.e., mode) after the TI, we need to instantaneously change the stiffnesses in a uniform manner (Fig.\ref{fig:methodology}):
\newline
\textbf{i. No ground stiffnesses:} the necessary conditions are 
$K_l^{(S2)}=\epsilon_l~K_l^{(S1)}$ where $l=1:n$ and $\epsilon_1=\epsilon_2=\dots=\epsilon_n=\epsilon$. The conditions guarantee that the eigenvalues of both states correlate such that:  $\omega_{l}^{(S2)}(\kappa)=\sqrt{\epsilon}~\omega_{l}^{(S1)}(\kappa)$ (Fig.\ref{fig:methodology_1} and Fig.\ref{fig:methodology_4}).
\newline
\textbf{ii. Adding equal ground stiffnesses:} if each mass is connected to the ground through $K_g$, additional necessary conditions: $K_g^{(S2)}=\lambda~K_g^{(S1)}$ and $\lambda=\epsilon_1=\epsilon_2=\dots=\epsilon_n$ need to be satisfied. Again, we establish the same relationships between the eigenvalues of both states (Fig.\ref{fig:methodology_2}).
\newline
\textbf{iii. Adding different ground stiffnesses:} we need to select initial ground stiffnesses proportional to the unit cell masses such that: $\frac{K_{g1}}{m_1}=\frac{K_{g2}}{m_2}=....=\frac{K_{gn}}{m_n}$. We also need to uniformly change the inter stiffnesses ($\epsilon_1=\epsilon_2=\dots=\epsilon_n=\epsilon$) and the ground stiffnesses ($\lambda_1=\lambda_2=\dots=\lambda_n=\lambda$). The conditions guarantee that the eigenvalues of both states correlate such that: $(\omega_l^{(S2)})^2-\epsilon~(\omega_l^{(S1)})^2=(\lambda-\epsilon)~\frac{K_{gl}}{m_l}$ (Fig.\ref{fig:methodology_3}).

Consequently, the orthogonality of the mode shapes leads to vanishing probability for all modes after the TI except a single mode such that: 
\begin{equation}
    \Psi_{i,j}=\phi_{S1}^{i}(\kappa_0)^\dag ~~\mathbf{M}~~\phi_{S2}^{j}(\kappa_0)~=~\begin{cases} \mbox{0,} & \mbox{if }~~i\ne j  \\ \mbox{1,} & \mbox{if}~~ i=j \end{cases} 
\end{equation}
which guarantees a single frequency appearance after the TI. 
\newline
On the other hand, violating these conditions for any of the considered cases leads to multi-mode conversion (multiple frequencies) after the temporal interface. It is worth noting that $\sum_{j=1}^n|\Psi_{i,j}|^2=1$. In some scenarios, $\Psi_{i,l} \approx0$, which means that mode $l$ has a negligible presence after the TI (Fig.\ref{fig:Diatomic} d-e). It is also worth noting that a system with negative $\epsilon$ makes a swap between modes, where the wave hops to a farther branch and/or  flip its group velocity sign (Fig.\ref{fig:focusing}).   
\newline
\section{Appendix B: Temporal Snell's law and Fresnel equations}
Consider a diatomic mass-spring system. Each unit cell contains two masses $m_1$ and $m_2$. The masses are connected to each other through inter stiffnesses $K_1$ and $K_2$ and to the ground through $K_{g1}$ and $K_{g2}$. The dispersion relation can be expressed as:
\begin{equation}
    \omega_{\substack{ac. \\ opt.}}^2(\kappa)=\frac{1}{2m_1m_2}\left ((m_1~\Tilde{K_1}+m_2~\Tilde{K_2})\mp \sqrt{S(\kappa)} \right)
\end{equation}
where: $\Tilde{K_1}=K_1+K_2+K_{g1}$, $\Tilde{K_2}=K_1+K_2+K_{g2}$ and $S (\kappa)=(m_1\Tilde{K_1}+m_2\Tilde{K_2})^2-4m_1m_2(\Tilde{K_1}\Tilde{K_2}-K_1^2-K_2^2-2K_1K_2cos(\kappa a))$. The group velocity can be calculated as:
\begin{equation}
    C_{\substack{ac. \\opt.}}(\kappa)=\frac{1}{\omega_{\substack{ac.\\opt.}}} \frac{\pm K_1K_2asin(\kappa a)}{\sqrt{S(\kappa)}}=\frac{\pm q}{\omega_{\substack{ac.\\opt.}}}
\end{equation}
Then, 
\begin{equation}
    |\frac{C_{j}^{(S2)}}{C_{i}^{(S1)}}|=|\frac{q^{(S2)}}{q^{(S1)}}~\frac{\omega_i^{(S1)}}{\omega_j^{(S2)}}|=|\frac{tan(\alpha _l)}{tan(\alpha_0)}|
\end{equation}
where $C_j^{(S2)}$ ($C_i^{(S1)}$) is the group velocity of a wave packet in $S_2$ ($S_1$) with a frequency $\omega_j^{(S2)}$ ($\omega_i^{(S1)}$), $q$ is a modification variable that depends on the system parameters and the incident wavenumber, and  $\alpha_l$ ($\alpha_0$) is the angle of the transmitted (incident) wave after (before) the TI. Equation A4 represents the modified Snell's law at the multiple choices temporal interfaces. 
The Bloch's form of the solution for the incident wave before the temporal interface is:
\begin{equation}
    u_n(t)=U_i~e^{i(n\kappa_0+\omega_0t)}~~~~~~~~~~~~~~~~~~~~~~~~~~~~~~~~~~~~~~~ t\le \tau
\end{equation}
where $U_i$ is the incident wave amplitude of $m_1$ or $m_2$.  The angular frequency $\omega_0$ and wavenumber $\kappa_0$ before the temporal interface satisfy the dispersion relation of state 1 (i.e., $S_1$). After the temporal interface, the general solution of the displacement can be expressed as a function of transmitted waves ($T_1$ and $T_2$) and reflected waves ($R_1$ and $R_2$) as: 
\begin{multline}
u_n(t)=U_{T1}~e^{i(n\kappa_0+\omega_1t)}+U_{T2}~e^{i(n\kappa_0-\omega_2t)}\\
+U_{R1}~e^{i(n\kappa_0-\omega_1t)}+U_{R2}~e^{i(n\kappa_0+\omega_2t)}~~~~t>\tau
\end{multline}
Where $U_{T1}$ and $U_{R1}$ are the amplitudes of the transmitted and reflected waves with frequency $\omega_1$ after the temporal interface. Similarly, $U_{T2}$ and $U_{R2}$ are the amplitudes of the transmitted and reflected waves with frequency $\omega_2$ after the temporal interface.    
To guarantee the continuity at the temporal interface, two conditions should be satisfied. The first constraint is the conservation in displacement:  $u_n|_{t=\tau^-}=u_n|_{t=\tau^+}$ where $\tau$ is the temporal interface time.   
\begin{multline}
U_i~e^{i\omega_0\tau}=U_{T1}~e^{i\omega_1\tau}+U_{T2}~e^{-i\omega_2\tau}+\\U_{R1}~e^{-i\omega_1\tau}+U_{R2}~e^{i\omega_2\tau}
\end{multline}
For simplicity, we can reduce the last equation to a simpler form:
\begin{multline}
    \Tilde{U}_i= \Tilde{U}_{T1}+\Tilde{U}_{T2}+\Tilde{U}_{R1}+\Tilde{U}_{R2}~~~~~~~~~~~~~~~~~
\end{multline}
The second constraint is the continuity of velocity: $\frac{\partial u_n(t)}{\partial t}|_{t=\tau^-}=\frac{\partial u_n(t)}{\partial t}|_{t=\tau^+}$.
\begin{multline}
    \omega_0\Tilde{U}_i= \omega_1\Tilde{U}_{T1}-\omega_2\Tilde{U}_{T2}-\omega_1\Tilde{U}_{R1}+\omega_2\Tilde{U}_{R2}~~~
\end{multline}
By using the overlap between modes, let $\Tilde{U}_{T1}=|\Psi_{i,1}|^2~\Tilde{U}_{T}$,  $\Tilde{U}_{T2}=|\Psi_{i,2}|^2~\Tilde{U}_{T}$, $\Tilde{U}_{R1}=|\Psi_{i,1}|^2~\Tilde{U}_{R}$ and $\Tilde{U}_{R2}=|\Psi_{i,2}|^2~\Tilde{U}_{R}$ where $|\Psi_{i,1}|^2$ and $|\Psi_{i,2}|^2$  are the probabilities of $\omega_1$ and $\omega_2$ appearing after a wave with $\omega_0$ crosses a temporal interface. Then, the continuity of the displacement field becomes:  
\begin{multline}
   \Tilde{U}_i= (|\Psi_{i,1}|^2+|\Psi_{i,2}|^2)~\Tilde{U}_{T}+(|\Psi_{i,1}|^2+|\Psi_{i,2}|^2)~\Tilde{U}_{R}\\=\Tilde{U}_{T}+\Tilde{U}_{R}~~~~~~~~~~~~~~~~~~~~~~~~~~~~~~~~~~~~~~~~~~~~~~~~~~~~~~~~~~~
\end{multline}
while the continuity of the velocity field becomes: 
\begin{multline}
    \omega_0\Tilde{U}_i= (|\Psi_{i,1}|^2~\omega_1-|\Psi_{i,2}|^2~\omega_2)\Tilde{U}_{T}+\\(-|\Psi_{i,1}|^2~\omega_1+|\Psi_{i,2}|^2~\omega_2)\Tilde{U}_{R}
\end{multline}
By solving both conditions, the amplitude of each wave after the temporal interface can be expressed as a function of the incident wave amplitude as:
\begin{equation}
    \frac{\Tilde{U}_{T1}}{\Tilde{U}_i}=\frac{1}{2}|\Psi_{i,1}|^2~\left( 1+\frac{\omega_0}{\sum_{j=1}^2~|\Psi_{i,j}|^2~\omega_j}\right)
\end{equation}
\begin{equation}
    \frac{\Tilde{U}_{T2}}{\Tilde{U}_i}=\frac{1}{2}|\Psi_{i,2}|^2~\left( 1+\frac{\omega_0}{\sum_{j=1}^2~|\Psi_{i,j}|^2~\omega_j}\right)
\end{equation}

\begin{equation}
    \frac{\Tilde{U}_{R1}}{\Tilde{U}_i}=\frac{1}{2}|\Psi_{i,1}|^2~\left( 1-\frac{\omega_0}{\sum_{j=1}^2~|\Psi_{i,j}|^2~\omega_j}\right)
\end{equation}

\begin{equation}
    \frac{\Tilde{U}_{R2}}{\Tilde{U}_i}=\frac{1}{2}|\Psi_{i,2}|^2~\left( 1-\frac{\omega_0}{\sum_{j=1}^2~|\Psi_{i,j}|^2~\omega_j}\right)
\end{equation}
These equations are the temporal analog of Fresnel equations
at a temporal interfaces with multiple choices (Fig.\ref{fig:Example_Fernsel equations}).
\section{Appendix C: Experimental methods}

 We utilize magnetic lattices to study temporal interfaces in polyatomic discrete structures (Fig.\ref{fig:setup}). Magnetic particles are free to hover over an air bearing table to minimize the friction. Each disk is surrounded by four fixed magnets confined within a stationary boundary. Each disk has a permanent dipole magnet and can move in two decoupled modes: longitudinal or shear. We excite the structure using an electromagnet (excitation source) connected to a signal generator. We place an electromagnet (tuning source) on each side of the unit cell to modulate the ground stiffness over time. Tuning sources are connected to a power supply to change the supplied voltage and hence, the effective ground stiffness. We capture the motion of each disk through high speed cameras. We analyze the captured photos using (DICe) to find the time response of each disk.
 We calibrate the electromagnets using monoatomic structure (m=0.34 g) (Fig.\ref{fig:setup}-a). We turn the electromagnets on with a given voltage, excite the structure with a chirp signal, and measure the transmission region experimentally ($f_{min.}$ and $f_{max.}$). For longitudinal motion, both the inter stiffness and the boundary's ground stiffness are positive  (Fig.\ref{fig:calibration long}-a), where $f_{min.}=\sqrt{\frac{K_g}{m}}$ and $f_{max.}=\sqrt{\frac{4K_{in}+K_g}{m}}$. Using these expressions, we calculate the effective ground stiffness corresponding to the supplied voltage (Fig.\ref{fig:calibration long}-b) and the corresponding transmission region (Fig.\ref{fig:calibration long}-c). We show the experimental results of three different values of voltage in Figure \ref{fig:calibration long}-(d-f). For shear motion, the inter stiffness is negative while the ground stiffness is positive (Fig.\ref{fig:calibration shear}-a). In such case, $f_{max.}=\sqrt{\frac{K_g}{m}}$ and $f_{min.}=\sqrt{\frac{4K_{in}+K_g}{m}}$. We measure the effective ground stiffness (Fig.\ref{fig:calibration shear}-b) and the transmission region (Fig.\ref{fig:calibration shear}-c) corresponding to three different values of voltage (Fig.\ref{fig:calibration shear}-(d-f)).
 \section{Appendix D: Topology detection from temporal interfaces}
 The boundaries of the topological phase transition are special states where the band gap closes at the degenerate point of frequency. To obtain these boundaries, we find the necessary conditions to make $S(\kappa)=0$. The first condition is $m_2~\Tilde{K_1}=m_1~\Tilde{K_2}$ which represents the protection of the chiral symmetry. The second condition depends on the wavenumber;   at $\kappa=0$ or $2\pi$, $K_1=-K_2$  while at $\kappa=\frac{\pi}{2}$, $K_1=K_2$ are the necessary conditions (Fig.\ref{fig:topology_SI}). The winding number is calculated as:
 \begin{equation}
    \nu=\int_{-\pi}^{\pi} \frac{1}{4\pi i}~tr[\mathbf{\sigma_3}~\mathbf{C}(\kappa)^{-1}~\partial_{\kappa}\mathbf {C}(\kappa)] d\kappa 
\end{equation}
where $\mathbf{\sigma_3}$ is the third pauli matrix and $\mathbf{C} (\kappa)$ is the chiral matrix obtained by eliminating the diagonal elements from the dynamic matrix.
\newline
The revival amplitude ``Loschmidt amplitude" which quantifies the deviation of the time-evolved state from 
the initial condition can be calculated as:
\begin{equation}
    g(\kappa,t)=\sum_{j=1}^2~ |\Psi_{1,j} (\kappa)|^2~exp(-i\omega^{(S2)}_j(\kappa)~t)
\end{equation}
At the critical wavenumber $\kappa_{DQPT}$, the revival amplitude vanishes periodically at critical times  $t_{critical}=(M-\frac{1}{2})~T_p$ where $M=1,2,3,\dots$, and $T_p$ is the period. By contrast, the revival amplitude
does not vanish for the initial state not at the critical wavenumber (Fig.\ref{fig:topology_SI2}). Moreover, we calculate the rate function $r(t)$ as follows:
\begin{equation}
    r(t)=-(\frac{1}{N_\kappa})~ln ~(\prod_\kappa |g~(\kappa,t)|^2~)
\end{equation}
where $N_\kappa$ is the number of wavenumbers considered in BZ (i.e., wavenumber discretizations). At critical times, nonanalytic cusps or kinks appear in the rate function $r(t)$.

\newpage 

\beginsupplement 
\newpage
\widetext
\newpage

\vspace{3pt}
{\huge \centering{\textbf{Supplementary Information:\\}} \vspace{9pt}
\huge{Topological mode conservation and conversion in phononic crystals with temporal interfaces}}\\

\large{Mahmoud Samak} and {Osama R. Bilal}

{School of Mechanical, Aerospace, and Manufacturing Engineering, University of Connecticut, Storrs, CT, 06269, USA.\\}

 \textbf{E-mail}: osama.bilal@uconn.edu\\
 
 \large\textbf{This PDF file includes:}
 \begin{itemize}
     \item Supporting text
     \item Supporting Figs. S1 to S11
 \end{itemize}

\newpage
\section{\large{Experimental methods}}

\begin{figure*}[!b] 
\includegraphics [width =  \columnwidth]{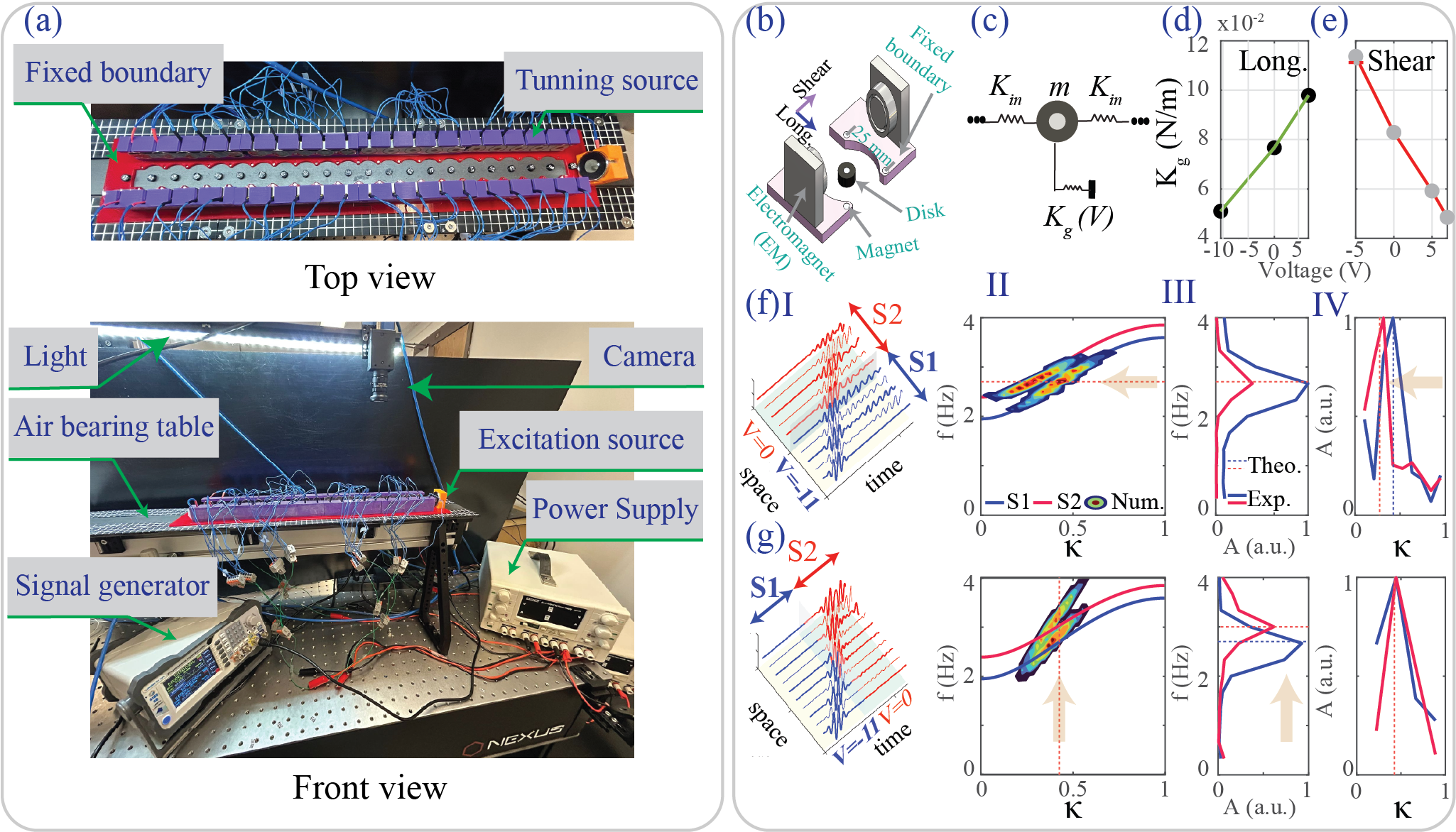} 
\caption{\label{fig:setup}\textbf{Spatial V.S. temporal interfaces in monoatomic lattices.} (a) Experimental setup: Magnetic particles confined in a fixed boundary are hovering over an air bearing table to minimize the friction. Each disk is centered between two  electromagnets (purble) to change its ground stiffness through controlling the supplied voltage from a power supply. An electromagnet (orange) connected to a signal generator is used to excite the structure. The motion of each disk is captured using an aye-bird camera . (b) Physical unit cell. (c) Mathematical model. Experimental measured ground stiffness as a function of the supplied voltage for (d) longitudinal and (e) shear motion. (f) Spatial interface and (g) temporal interface between two states -11 V and 0 V. For each interface we show (I) wave packet propagation. (II) analytical dispersions and numerically calculated 2D FFT.  Experimentally observed (III) frequencies and (IV) wavenumbers before and after the interface.   }
\end{figure*}
\begin{figure*} 
\includegraphics [scale= 0.85]{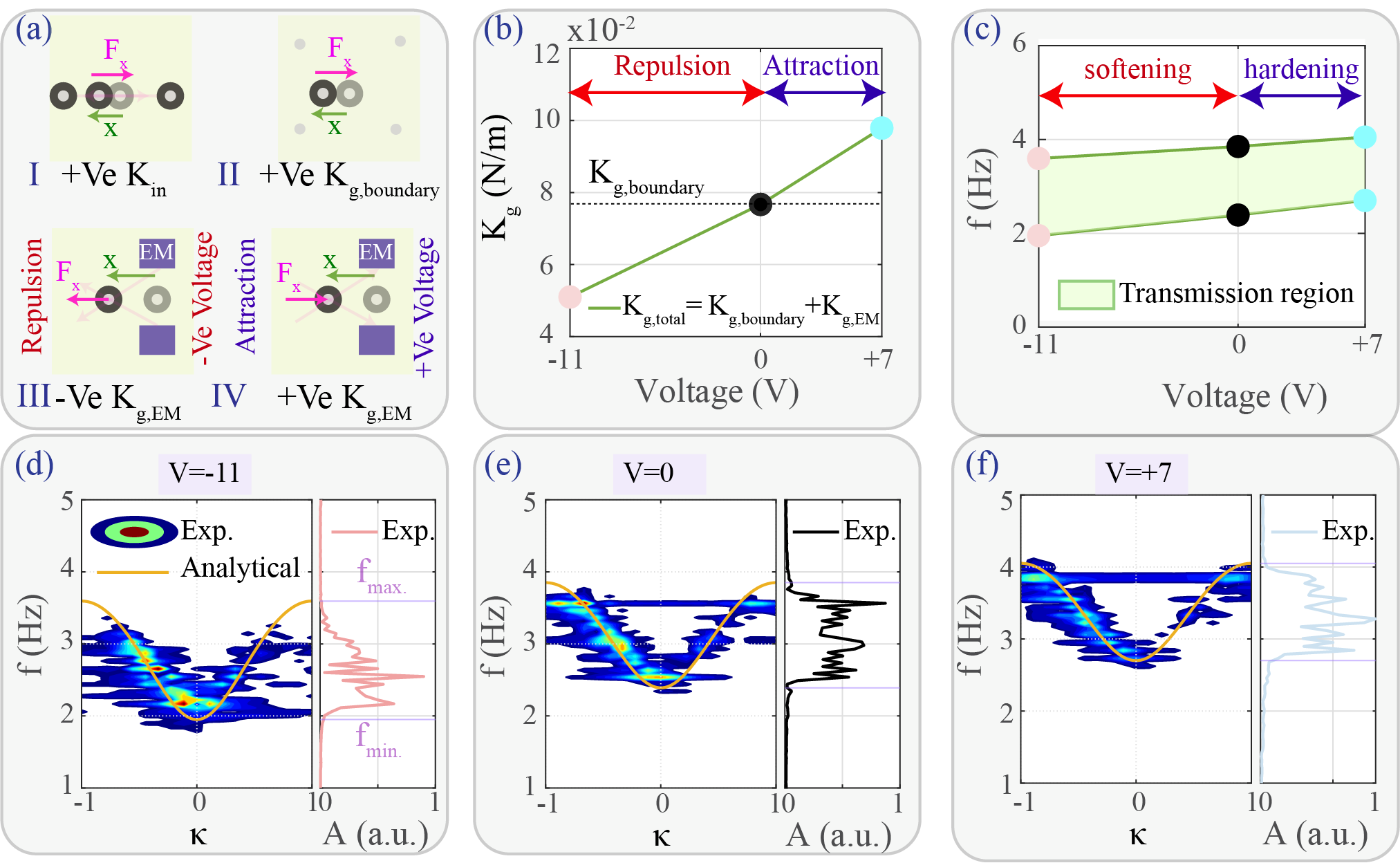} 

\caption{\label{fig:calibration long}\textbf{Experimental measurement of longitudinal ground stiffness.} (a) The horizontal force direction due to a small perturbation from the equilibrium position on a disk between (I) neighbor disks, (II) boundary magnets, (III) two electromagnets with negative voltage and (IV) two electromagnets with positive voltage. (b) The total ground stiffness in the longitudinal direction as a function of the supplied voltage. (c) The transmission region of the longitudinal motion as a function of the supplied voltage. Analytical dispersion curve over the experimental 2D FFT (left) and the experimental measured transmission region (right) for: (d) V=-11, (e) V=0 and (f) V=+7.     }
\end{figure*}

\begin{figure*} 
\includegraphics
[scale= 0.85]{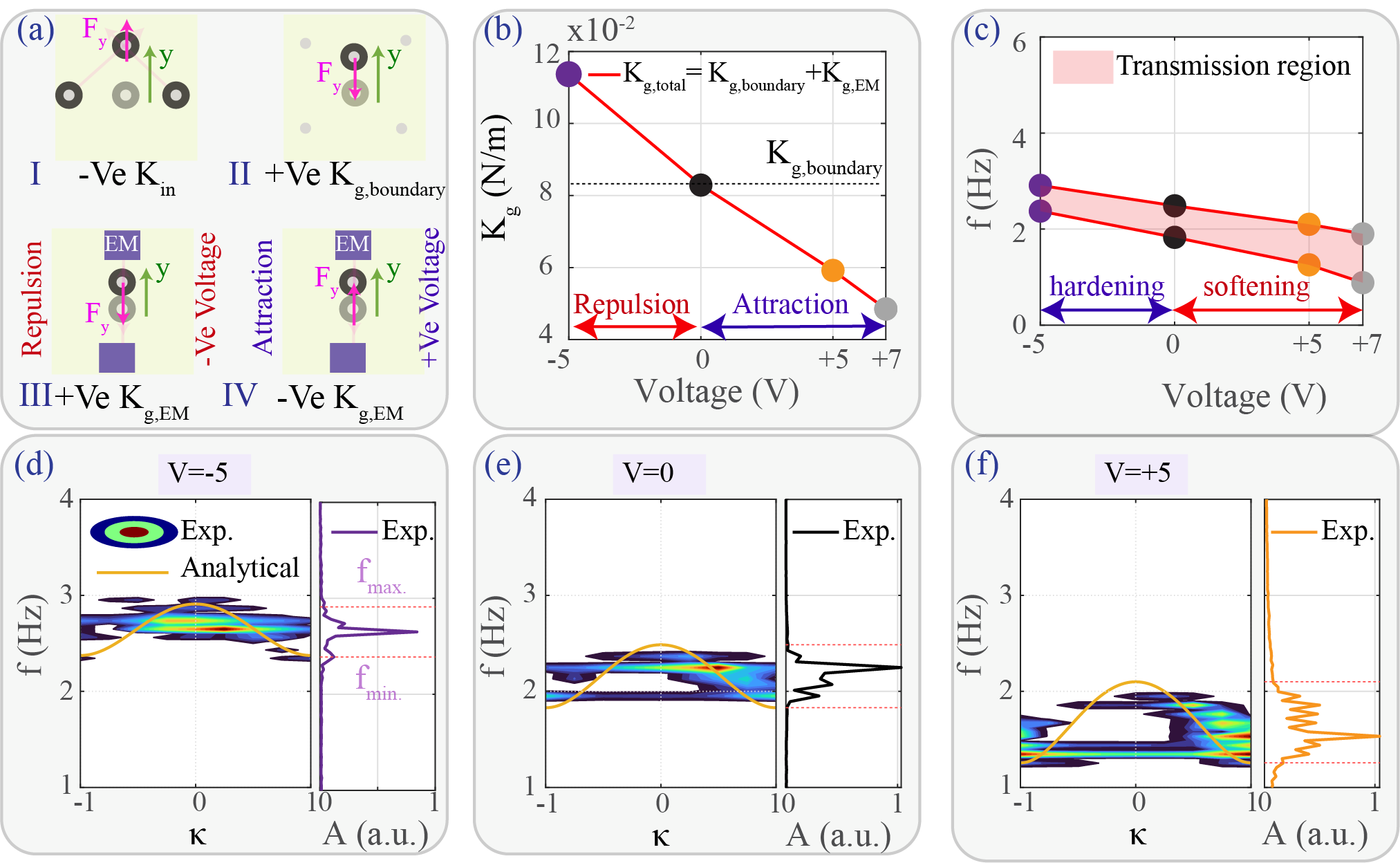} 
\caption{\label{fig:calibration shear}\textbf{Experimental measurement of shear ground stiffness.}  (a) The vertical force direction due to a small perturbation from the equilibrium position on a disk between (I) neighbor disks, (II) boundary magnets, (III) two electromagnets with negative voltage and (IV) two electromagnets with positive voltage. (b) The total ground stiffness in the shear direction as a function of the supplied voltage. (c) The transmission region of the shear motion as a function of the supplied voltage. Analytical dispersion curve over the experimental 2D FFT (left) and the experimental measured transmission region (right) for: (d) V=-5, (e) V=0 and (f) V=+5. }
\end{figure*}

\paragraph{Experimental setup} We utilize magnetic lattices to study temporal interfaces in polyatomic discrete structures (Fig.\ref{fig:setup}). Magnetic particles are free to hover over an air bearing table to minimize the friction. Each disk is surrounded by four fixed magnets confined within a stationary boundary. Each disk has a permanent dipole magnet and can move in two decoupled modes: longitudinal or shear. We excite the structure using an electromagnet (excitation source) connected to a signal generator. We place an electromagnet (tuning source) on each side of the unit cell to modulate the ground stiffness over time. Tuning sources are connected to a power supply to change the supplied voltage and hence, the effective ground stiffness. We capture the motion of each disk through high speed cameras. We analyze captured photos using digital image correlation engine (DICe) to find the time response of each disk.

\paragraph{Electromagnet calibration} We calibrate the electromagnets using a monoatomic structure (Fig.\ref{fig:setup}-b). Each unit cell contains two electromagnets at the center of the unit cell. The electromagnets are connected to a power supply to control the supplied voltage. First, we turn the electromagnets on with a specific voltage. Second, we excite the structure with a chirp signal. Third, we measure the transmission region experimentally ($f_{min.}$ and $f_{max.}$). For longitudinal motion, the inter stiffness and the ground stiffness from boundary are positive (Fig.\ref{fig:calibration long}-a). In such case, $f_{min.}=\sqrt{\frac{K_g}{m}}$ and $f_{max.}=\sqrt{\frac{4K_{in}+K_g}{m}}$. Using these expressions, we calculate the effective ground stiffness corresponding to the supplied voltage (Fig.\ref{fig:calibration long}-b) and the corresponding transmission region (Fig.\ref{fig:calibration long}-c). We show the experimental results of three voltages (Fig.\ref{fig:calibration long}-(d-f)). For shear motion, the inter stiffness is negative while the ground stiffness is positive (Fig.\ref{fig:calibration shear}-a). In such case, $f_{max.}=\sqrt{\frac{K_g}{m}}$ and $f_{min.}=\sqrt{\frac{4K_{in}+K_g}{m}}$. We measure the effective ground stiffness (Fig.\ref{fig:calibration shear}-b) and the transmission region (Fig.\ref{fig:calibration shear}-c) corresponding to three different voltages (Fig.\ref{fig:calibration shear}-(d-f)).

\paragraph{Spatial V.S. temporal interfaces}To elucidate the differences between a spatial and a temporal interface, for a monoatomic lattice, we consider a finite structure composed of 20 identical unit cells. For the spatial interface (Fig. \ref{fig:setup}-f), the first half of the structure (i.e., first ten unit cells) have a supplied voltage of $-11V$ to each electromagnet, while for the second half, the electromagnets are turned off (i.e., $0V$) (Fig.\ref{fig:setup}-f(I)). We excite the structure in the longitudinal direction with a wave packet centered at $f=2.7$Hz. As the wave packet crosses the spatial interface, the frequency remains the same (i.e.,  conserved), while the wavenumber changes (i..e, converted). We observe the frequency conservation and the wavenumber conversion both in the numerical 2D FFTs of the lattice displacements (Fig.\ref{fig:setup}-f(II)), and the experimental FFTs of a single disk before and after the interface (Fig.\ref{fig:setup}-f(III-IV)). For the temporal interface, the entire structure (i.e., all 20 unit cells) is  supplied with a voltage of $-11V$ to each electromagnet at the beginning of the excitation, once the wave packet reaches the middle of the structure, the electromagnets are turned off (i.e., $0V$) causing a sudden change in the ground stiffness (Fig.\ref{fig:setup}-g). In contrast to the spatial interface, we observe a clear preservation of the wavenumber and a conversion of frequency both numerically (Fig.\ref{fig:setup}-g(II)) and experimentally (Fig.\ref{fig:setup}-g(III-IV)).

\section{\large{Mode shapes conservation in polyatomic lattices}}
\begin{figure*} 
\includegraphics
[scale= 1]{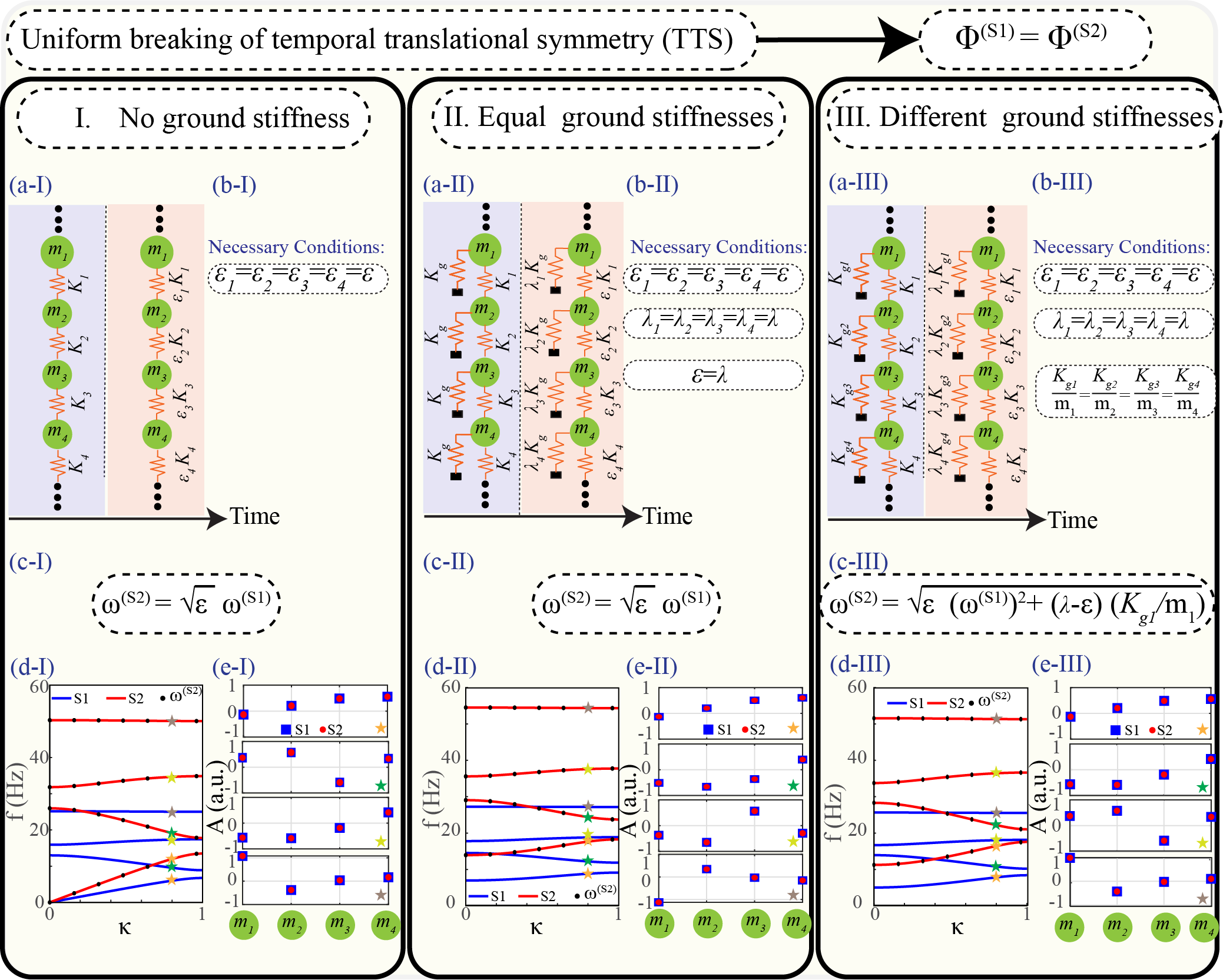} 
\caption{\label{fig:methodology}\textbf{Design methodology for uniform breaking of temporal translational symmetry (TTS).} poly-atomic structure with four masses per unit-cell ( $m_1=\frac{m_2}{2}=\frac{m_3}{3}=\frac{m_4}{4}=0.1$ Kg) and initially $K_1=K_2=K_3=K_4=1000$ N/m is subjected to a temporal change to S2. (I) No ground stiffnesses with $\epsilon_l$=4 where $l=1:4$. (II) Equal ground stiffnesses are added to each mass ($K_g$=500 N/m) with $\epsilon_l=\lambda_l=4$ where $l=1:4$. (III) Different ground stiffnesses are added such that $\frac{K_{gl}}{m_l}$=1000 $(rad/s)^2$  with $\epsilon_l$=4 and $\lambda_l$=5 where $l=1:4$. For each case, we show (a) mathematical model. (b) corresponding  necessary conditions to uniform breaking TTS. (c) relationship between frequencies in $S_2$ and $S_1$ at the same wavenumber. (d) dispersion branches for $S_1$ (blue), $S_2$ (red), and $S_2$ from equations in (c) (black dots). (e) mode shapes of $S_1$ (blue squares) and $S_2$ (red circles) at $\kappa=0.8\frac{\pi}{a}$ for the first branch (top) to the forth branch (bottom). The matching between mode shapes show mode conservation theory as a result of uniform breaking of TTS.        }
\end{figure*}

At temporal interfaces, the propagating wave conserves the wavenumber while converting the frequency. In monoatomic lattices, a single frequency corresponds to each wavenumber, which allows only one solution. However, in polyatomic lattices, multiple frequencies correspond to the same wavenumber, which opens the door to many solutions. We show that the probability of each frequency (mode shape) depends on the degree of matching between the mode shape at the incident point ($\kappa_0$ , $\omega_0$) and the new state point(s) ($\kappa_0$ , $\omega_l$). Such degree of matching can be calculated as the overlap  between two modes: $\Psi_{i,j}=\mathbf{\phi_{S1}^i}(\kappa_0)^\dag ~~\mathbf{M}~~ \mathbf{\phi_{S2}^j} (\kappa_0)$, where $\mathbf{M}$ is the mass matrix, \textbf{ $\phi_{S1}^i$} is the mode shape on the dispersion branch $i$ at state ($S_1$) before the TI and \textbf{$\phi_{S2}^j$} is the mode shape at the dispersion branch $j$ at state ($S_2$) after the TI, $\kappa_0$ is the wavenumber corresponding to the excitation frequency,  $(.)^\dag$ is the complex conjugate transpose and $\sum_{j=1}^n|\Psi_{i,j}|^2=1$. According to such equation, if \textbf{ $\phi_{S1}^i$}=\textbf{$\phi_{S2}^j$} ``mode conservation", $\Psi_{i,j}=1$ which means that an incident wave on branch $i$ of $S_1$ before TI will transfer completely to branch $j$ of $S2$ after TI (i.e., single frequency appears after TI). To design for mode conservation scenario, we need to carefully tune the system's parameters in $S2$ such that dispersion branches have shift in eigenvalues while fixing the eigenvectors. To satisfy such constraint, we need to break temporal translational symmetry (TTS) in a uniform way. In a general polyatomic lattice which has $n$ masses, $n$ inter stiffnesses and $n$ ground stiffnesses, $K_l^{(S2)}=\epsilon_l~K_l^{(S1)}$ and $K_{gl}^{(S2)}=\lambda_l~K_{gl}^{(S1)}$ where $l=1:n$. Necessary conditions for uniform breaking TTS become:
\newline
\textbf{Case i. no ground stiffness:} if $\epsilon_1=\epsilon_2=....=\epsilon_n=\epsilon$, then $\omega_{l}^{(S2) }(\kappa)=\sqrt{\epsilon}~\omega_{l}^{(S1)}(\kappa)$ and    $\phi_{S1}^{l}(\kappa)=\phi^{l}_{S2}(\kappa)$.
\newline
\textbf{Case ii. equal ground stiffness:} if $\epsilon_1=\epsilon_2=....=\epsilon_n=\epsilon$, $\lambda_1=\lambda_2=....=\lambda_n=\lambda$, $\epsilon=\lambda$ then $\omega_{l}^{(S2) }(\kappa)=\sqrt{\epsilon}~\omega_{l}^{(S1)}(\kappa)$ and    $\phi_{S1}^{l}(\kappa)=\phi^{l}_{S2}(\kappa)$.
\newline
\textbf{Case iii. different ground stiffness:} if $\epsilon_1=\epsilon_2=....=\epsilon_n=\epsilon$, $\lambda_1=\lambda_2=....=\lambda_n=\lambda$, and $\frac{K_{g1}}{m_1}=\frac{K_{g2}}{m_2}=.....=\frac{K_{gn}}{m_n}$ then $\omega_{l}^{(S2) }(\kappa)=\sqrt{\epsilon (\omega_{l}^{(S1)}(\kappa))^2+(\lambda-\epsilon)\frac{K_{gl}}{m_l}}~$ and    $\phi_{S1}^{l}(\kappa)=\phi^{l}_{S2}(\kappa)$. 
\newline
To summarize these cases, we consider a polyatomic lattice with four masses per unit cell with $m_1=\frac{m_2}{2}=\frac{m_3}{3}=\frac{m_4}{4}=0.1$ Kg and initially $K_1=K_2=K_3=K_4=1000$ N/m  (Fig.\ref{fig:methodology}). We first consider a lattice with no ground stiffness with $\epsilon_l=4$ where $l=1:4$ (Fig.\ref{fig:methodology} (a-I)). This scenario establishes a relationship between eigenvalues of both states such that $\omega_{l}^{(S2) }(\kappa)=\sqrt{4}~\omega_{l}^{(S1)}(\kappa)$ (Fig.\ref{fig:methodology} (c-I)). To validate such hypothesis, we plot the dispersion curves of $S_1$ (blue), $S_2$ (red) through solving the eigenvalue problem and $S2$ (black dots) through the equation $\omega_{l}^{(S2) }(\kappa)=2~\omega_{l}^{(S1)}(\kappa)$ (Fig.\ref{fig:methodology} (d-I)). The overlay between red lines and black dots proves the hypothesis. More interestingly, the mode shapes corresponding to the same wavenumbers are identical such that $\phi_{S1}^{l}(\kappa)=\phi^{l}_{S2}(\kappa)$ (Fig.\ref{fig:methodology} (e-I)). We further consider the same lattice with adding equal ground stiffness to each mass $K_g=500$ N/m (Fig.\ref{fig:methodology} (a-II)). We change the structure from $S1$ to $S2$ through $\epsilon=\lambda=4$ (Fig.\ref{fig:methodology} (b-II)). This scenario establishes a relationship between the eigenvalues of both states such that $\omega_{l}^{(S2) }(\kappa)=\sqrt{4}~\omega_{l}^{(S1)}(\kappa)$ (Fig.\ref{fig:methodology} (c-II)). We follow the same steps to prove our methodology for this case (Fig \ref{fig:methodology} (d-II)-(e-II)). Furthermore, we consider the same lattice with different ground stiffnesses connected to different masses such that $\frac{K_{gl}}{m_l}=1000 (rad/s)^2$ (Fig.\ref{fig:methodology} (a-III)). We change the structure from $S_1$ to $S_2$ through $\epsilon=4$ and $\lambda=5$ (Fig.\ref{fig:methodology} (c-III)). In this case, we establish a relationship between the eigenvalues of both states given by $\omega_{l}^{(S2) }(\kappa)=\sqrt{4 (\omega_{l}^{(S1)}(\kappa))^2+1000}$ (Fig.\ref{fig:methodology} (c-III)). By following the same steps, we can show the validity of such relationship and mode-shape conservation (Fig.\ref{fig:methodology} (d-III)-(e-III)). Our finding presents a design methodology to tune the system parameters in $S2$ to conserve mode shapes and hence obtain a single frequency after a temporal interface.           
\section{\large{Mode shapes conservation V.S. conversion in poly atomic lattices}}
\begin{figure*} 
\includegraphics
[scale= 0.85]{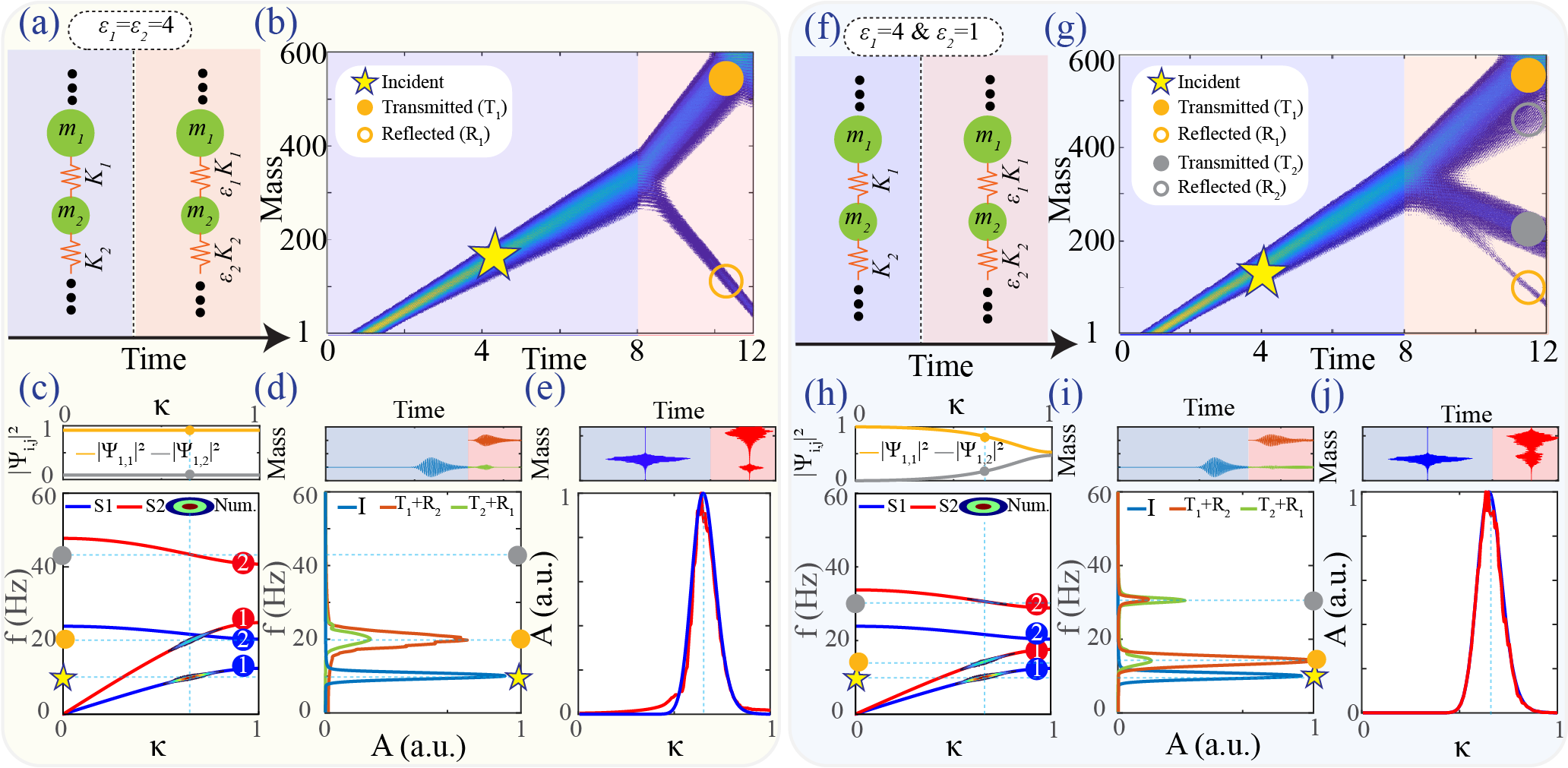} 
\caption{\label{fig:methodology_1}\textbf{Design methodology (no ground stiffnesses).} Diatomic structure with $m_1=0.5~m_2=0.1$ Kg and initially $K_1=0.5~K_2=500$ N/m is subjected to a temporal change to S2 (a) uniform change in stiffnesses ($\epsilon_1$=$\epsilon_2$=4) (b) Incident wave packet at 10 Hz is converted to a single frequency after TI. (c) (top) the probability of each mode of S2 after TI (bottom) analytical dispersion curves for each state with 2D FFT overlay. (d) (top) three time signals of mass 200 ($t<\tau$), 400 ($t>\tau$) and 200 ($t>\tau$) to represent incident, forward and backward waves. (bottom) FFT of the three signals in the top figure. (e) (top) two spatial profiles at $t=6$ s and $t=10$ s. (bottom) FFT of the two spatial profiles in the top figure. (f) non uniform change in stiffnesses ($\epsilon_1$=4 and $\epsilon_1$=1). (g) Incident wave packet at 10 Hz is converted to two frequencies after TI. (h) (top) the probability of each mode of S2 after TI (bottom) analytical dispersion curves for each state with 2D FFT overlay. (i) (top) three time signals of mass 200 ($t<\tau$), 400 ($t>\tau$) and 200 ($t>\tau$) to represent incident, forward and backward waves. (bottom) FFT of the three signals in the top figure. (j) (top) two spatial profiles at $t=6$ s and $t=10$ s. (bottom) FFT of the two spatial profiles in the top figure.      }
\end{figure*}

Next, we consider our design methodology with temporal interfaces showing either (i) a single frequency or (ii) multiple frequencies after breaking time translational symmetry. To show case (i), we consider a diatomic structure with $m_1=0.5~m_2=0.1$ Kg and initially $K_1=0.5~K_2=500$ N/m. First, we consider a uniform change in stiffnesses   ($\epsilon_1$=$\epsilon_2$=4) (Fig.\ref{fig:methodology_1} -a). We excite the first mass in a finite structure that contains 600 masses with a wave packet centered at 10 Hz (on branch 1 of $S_1$). Once the wave packet reaches the middle of the finite structure, we change the parameters of the structure from $S_1$ to $S_2$. A single pair of waves propagate after the temporal interface: transmitted wave to forward and reflected wave to backward (Fig.\ref{fig:methodology_1}-b). We calculate the probability of each mode of S2 after TI :$|\Psi_{1,1}|^2=1$ and $|\Psi_{1,2}|^2=0$. These calculated values show mode conservation between both states (Fig.\ref{fig:methodology_1}-c(top)). We plot the analytical dispersion curves for each state with an overlay of the 2D FFT of the spatiotemporal results (Fig.\ref{fig:methodology_1}-c(bottom)). Moreover, we pick three time signals of masses 200 ($t<8 ~sec$), 400 ($t> 8~sec$) and 200 ($t> 8~sec$) to represent incident, forward, and backward waves (Fig.\ref{fig:methodology_1}-d(top)). Then, we calculate the FFT of the three signals to find the frequencies present within the structure (Fig.\ref{fig:methodology_1}-d(bottom)). As a result of the uniform breaking of TTS, the incident wave (10 Hz) is converted to a single frequency after the temporal interface(20 Hz).  To show conservation of the wavenumber, we consider two spatial profiles at $t=6$ s and $t=10$ s (Fig.\ref{fig:methodology_1}-e(top)). We calculate the FFT of the two spatial profiles which show the wavenumber conservation (Fig.\ref{fig:methodology_1}-e(bottom)). Second, we consider a nonuniform change in stiffnesses ($\epsilon_1$=4 and $\epsilon_1$=1) (Fig.\ref{fig:methodology_1}-f). We excite the finite structure with a wave packet at 10 Hz ( on branch 1 of $S_1$) (Fig.\ref{fig:methodology_1}-g). Once the wave packet reaches the middle of the structure, a sudden change in stiffnesses converts the system to $S2$. Afterwards, two pairs of waves propagate through the structure: two waves forward (transmitted wave T1 and reflected wave R2) and two waves backward (transmitted wave T2 and reflected wave R1). We calculate the probability of each mode of S2 after the temporal interface such that $\sum_{j=1}^2|\Psi_{1,j}|^2=1$ with a significant contribution from both modes (Fig.\ref{fig:methodology_1}-h(top)). We plot the analytical dispersion curves of both states and overlay the 2D FFT of spatiotemporal results (Fig.\ref{fig:methodology_1}-h(bottom)). Similarly, we consider three time signals of masses 200 ($t<8~sec$), 400 ($t>8~sec$), and 200 ($t>8~sec$) to represent incident, forward and backward waves (Fig.\ref{fig:methodology_1}-i(top)).The FFT of the three signals show the frequency conversion from a single frequency before the temporal interface to two frequencies after the temporal (Fig.\ref{fig:methodology_1}-i(bottom)). We consider two spatial profiles at $t=6$ s and $t=10$ s (Fig.\ref{fig:methodology_1}-j(top)). The FFT of the two spatial profiles show the conservation of wavenumber (Fig.\ref{fig:methodology_1}-j(bottom)).

\begin{figure*} 
\includegraphics
[scale= 0.85]{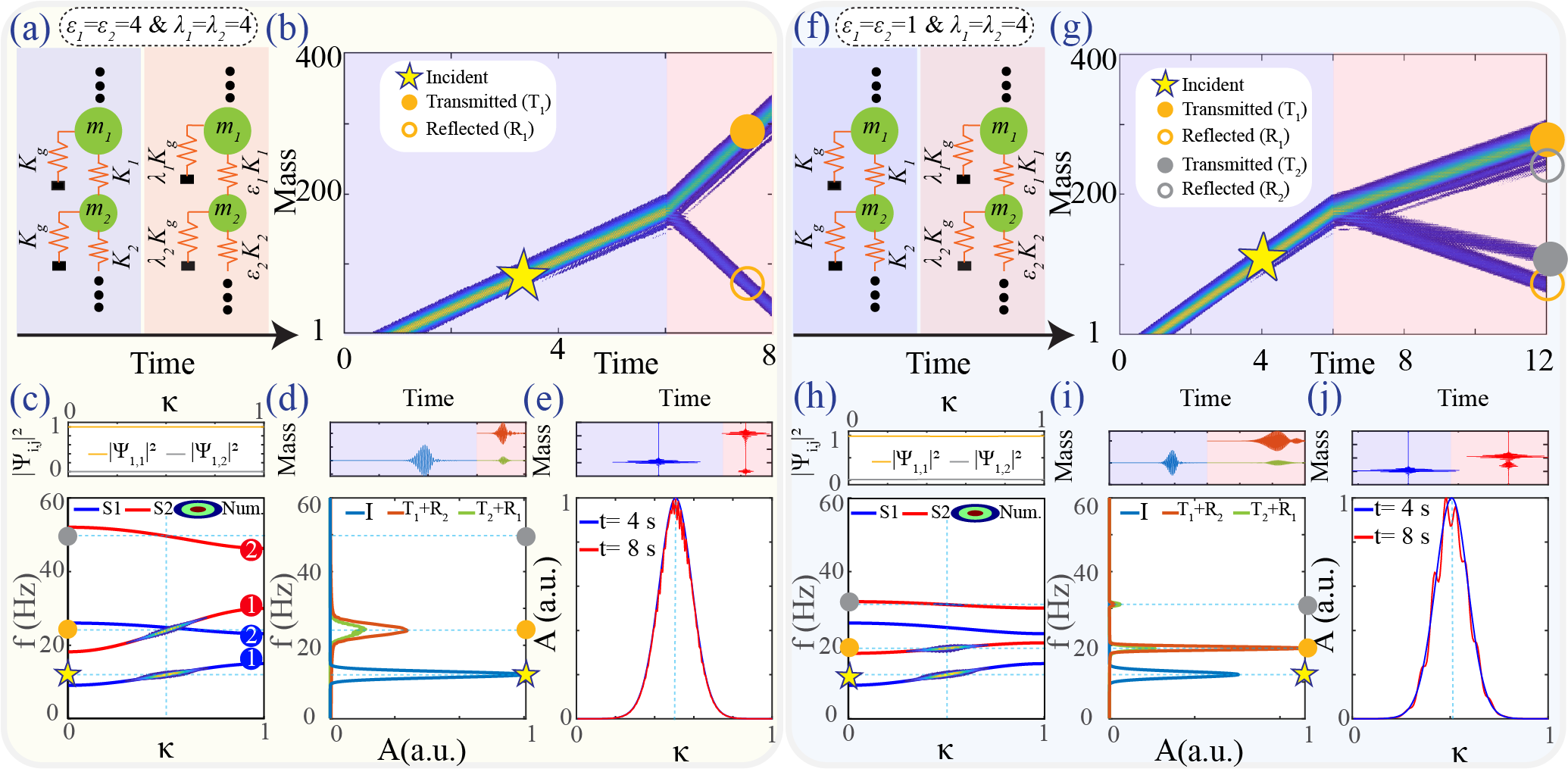} 
\caption{\label{fig:methodology_2}\textbf{Design methodology (equal ground stiffnesses).} Diatomic structure with $m_1=0.5~m_2=0.1$ Kg and initially $K_1=0.5~K_2=500$ N/m and $K_{g1}=K_{g2}=500$ N/m is subjected to a temporal change to S2 (a) uniform change in stiffnesses ($\epsilon_1$=$\epsilon_2$=$\lambda_1$=$\lambda_2$=4) (b) Incident wave packet at 12 Hz is converted to a single frequency after TI at 24 Hz. (c) (top) the probability of each mode of S2 after TI (bottom) analytical dispersion curves for each state with 2D FFT overlay. (d) (top) three time signals of mass 100 ($t<\tau$), 250 ($t>\tau$) and 100 ($t>\tau$) to represent incident, forward and backward waves. (bottom) FFT of the three signals in the top figure. (e) (top) two spatial profiles at $t=4$ s and $t=8$ s. (bottom) FFT of the two spatial profiles in the top figure. (f) non uniform change in stiffnesses ($\epsilon_1$=$\epsilon_2$=1 and $\lambda_1$=$\lambda_2$=4). (g) Incident wave packet at 12 Hz is converted to two frequencies after TI. (h) (top) the probability of each mode of S2 after TI (bottom) analytical dispersion curves for each state with 2D FFT overlay. (i) (top) three time signals of mass 100 ($t<\tau$), 250 ($t>\tau$) and 100 ($t>\tau$) to represent incident, forward and backward waves. (bottom) FFT of the three signals in the top figure. (j) (top) two spatial profiles at $t=4$ s and $t=8$ s. (bottom) FFT of the two spatial profiles in the top figure.      }
\end{figure*}

\begin{figure*} 
\includegraphics
[scale= 0.85]{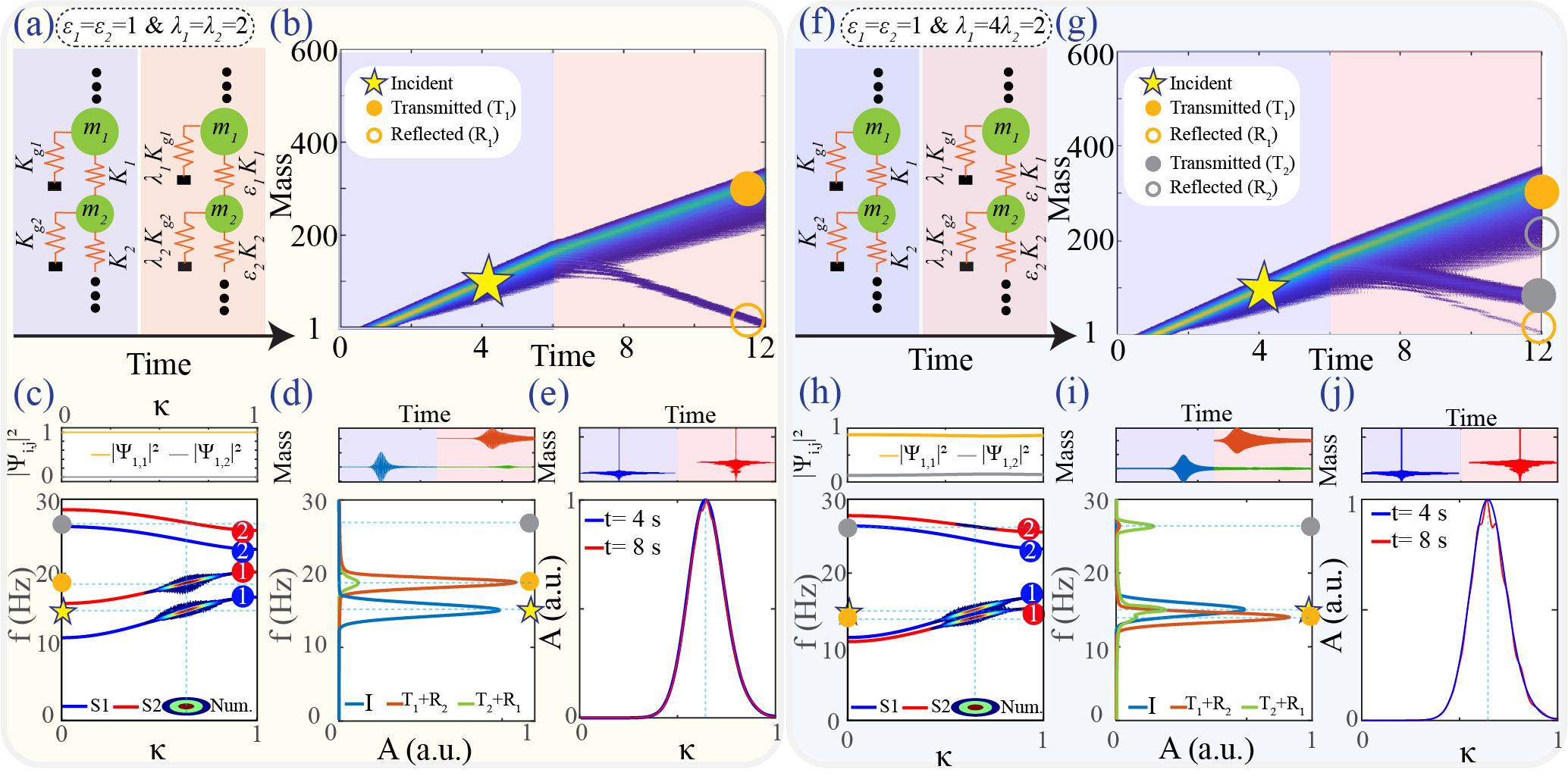} 
\caption{\label{fig:methodology_3}\textbf{Design methodology (different ground stiffnesses).} Diatomic structure with $m_1=0.5~m_2=0.1$ Kg and initially $K_1=0.5~K_2=500$ N/m and $K_{g1}=0.5~K_{g2}=500 $ N/m is subjected to a temporal change to S2 (a) uniform change in stiffnesses ($\epsilon_1$=$\epsilon_2$=1) and ($\lambda_1=\lambda_2=2$) (b) Incident wave packet at 15 Hz is converted to a single frequency after TI at 18.8 Hz. (c) (top) the probability of each mode of S2 after TI (bottom) analytical dispersion curves for each state with 2D FFT overlay. (d) (top) three time signals of mass 100 ($t<\tau$), 200 ($t>\tau$) and 100 ($t>\tau$) to represent incident, forward and backward waves. (bottom) FFT of the three signals in the top figure. (e) (top) two spatial profiles at $t=4$ s and $t=8$ s. (bottom) FFT of the two spatial profiles in the top figure. (f) non-uniform change in stiffnesses ($\epsilon_1=\epsilon_2=1$ and $\lambda_1=4~\lambda_2$=2). (g) Incident wave packet at 15 Hz is converted to two frequencies after TI. (h) (top) the probability of each mode of S2 after TI (bottom) analytical dispersion curves for each state with 2D FFT overlay. (i) (top) three time signals of mass 100 ($t<\tau$), 200 ($t>\tau$) and 100 ($t>\tau$) to represent incident, forward and backward waves. (bottom) FFT of the three signals in the top figure. (j) (top) two spatial profiles at $t=4$ s and $t=8$ s. (bottom) FFT of the two spatial profiles in the top figure.      }
\end{figure*}

\begin{figure*} 
\includegraphics
[scale= 0.85]{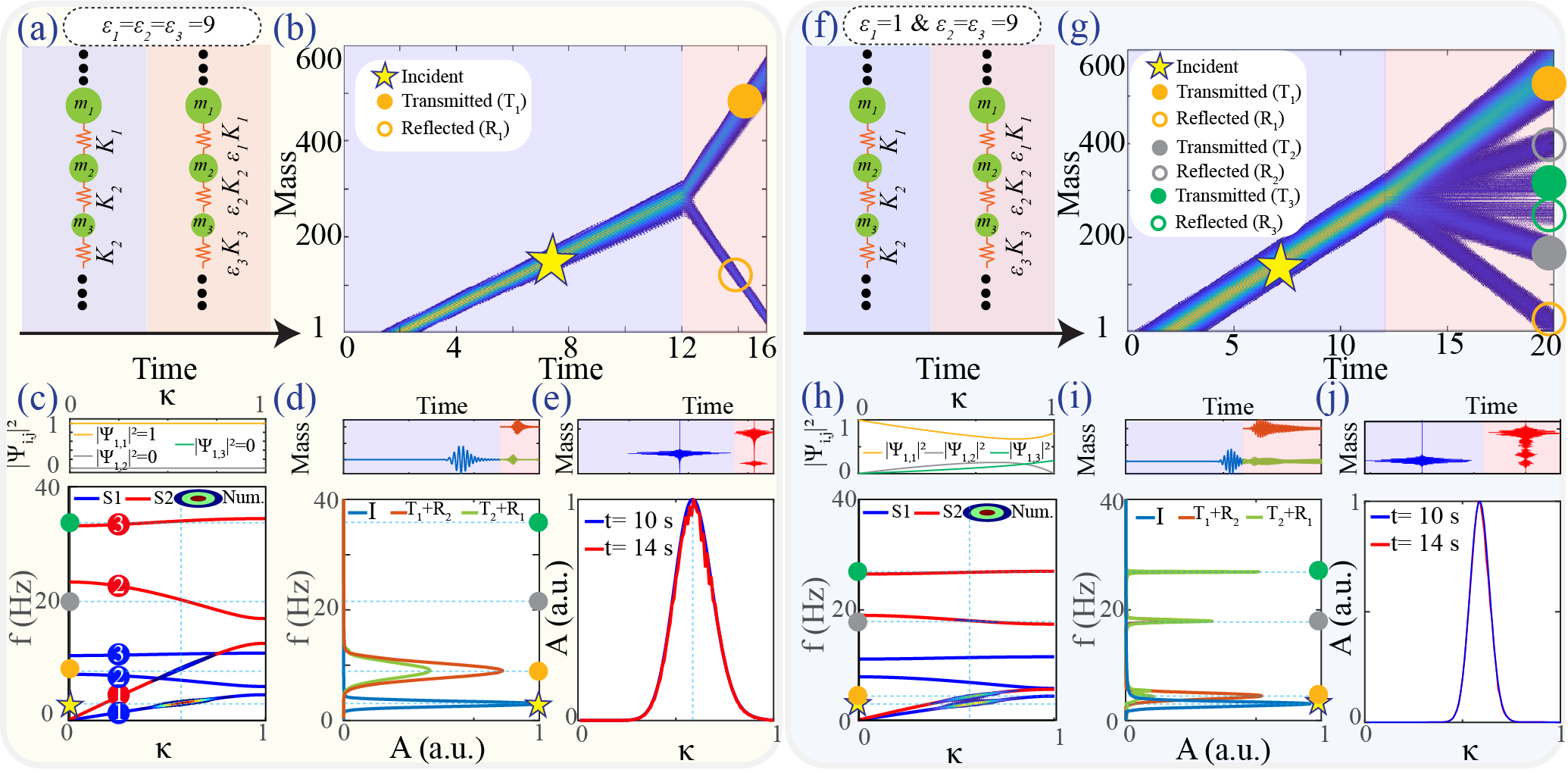} 
\caption{\label{fig:methodology_4}\textbf{Design methodology (no ground stiffness).} Triatomic structure with $m_1=0.1,~m_2=0.2,~$and $m_3=0.3$ Kg and initially $K_1=~K_2=~K_3=~200$ N/m is subjected to a temporal change to S2 (a) uniform change in stiffnesses ($\epsilon_1$=$\epsilon_2$=$\epsilon_3$=9) (b) Incident wave packet at 3 Hz is converted to a single frequency after TI at 9 Hz. (c) (top) the probability of each mode of S2 after TI (bottom) analytical dispersion curves for each state with 2D FFT overlay. (d) (top) three time signals of mass 200 ($t<\tau$), 400 ($t>\tau$) and 200 ($t>\tau$) to represent incident, forward and backward waves. (bottom) FFT of the three signals in the top figure. (e) (top) two spatial profiles at $t=10$ s and $t=14$ s. (bottom) FFT of the two spatial profiles in the top figure. (f) nonuniform change in stiffnesses ($\epsilon_1=1~$ and $\epsilon_2=\epsilon_3=9$. (g) Incident wave packet at 3 Hz is converted to three frequencies after TI. (h) (top) the probability of each mode of S2 after TI (bottom) analytical dispersion curves for each state with 2D FFT overlay. (i) (top) three time signals of mass 200 ($t<\tau$), 400 ($t>\tau$) and 200 ($t>\tau$) to represent incident, forward and backward waves. (bottom) FFT of the three signals in the top figure. (j) (top) two spatial profiles at $t=10$ s and $t=14$ s. (bottom) FFT of the two spatial profiles in the top figure.      }
\end{figure*}

\begin{figure*} 
\includegraphics
[scale= 1]{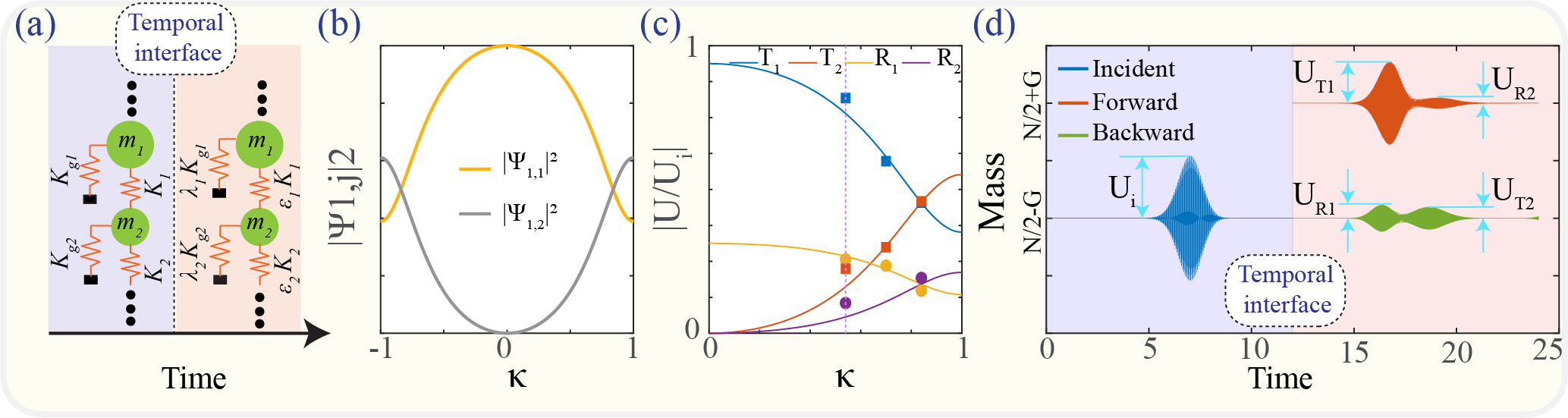} 
\caption{\label{fig:Example_Fernsel equations}\textbf{Validation of modified Fernsel equations in phononic lattices.} Diatomic structure with $m_1=0.5~m_2=0.1$ Kg and initially with $K_1=~0.5~K_2=~500$ N/m and $K_{g1}=0.5~K_{g2}=1000$ N/m. The structure is subjected to a temporal interface with nonuniform breaking of TTS ($\epsilon_1=8$,   $\epsilon_2=1, and ~ \lambda_1=\lambda_2=4$). (a) Mathematical model. (b) Analytical probability of each mode in $S2$ for an incident wave on mode 1 of $S_1$.  (c) Ratio between transmitted
( $U_{T1}$ and $U_{T2}$) or reflected ( $U_{R1}$ and $U_{R2}$) wave packet amplitude to the
incident wave packet amplitude ($U_i$) theoretically (lines) and numerically
(markers) to show Fresnel equations. Three wavenumbers are considered: ( 0.52, 0.7, 0.845)$\frac{\pi}{a}$.  (d) Time response of masses ($N/2-G$ for $t<\tau$), ($N/2+G$ for $t>\tau$) and ($N/2-G$ for $t>\tau$) to show incident, forward and backward waves (G is arbitrary integer) at $\kappa=0.52 \frac{\pi}{a}$. We calculate the amplitude of the incident wave packet $U_i$, first transmitted wave $U_{T1}$, first reflected wave $U_{R1}$, second transmitted wave $U_{T2}$, and second reflected wave $U_{R2}$.        }
\end{figure*}

In a similar way, we show an example of case (ii) and case (iii) in Figs.\ref{fig:methodology_2} and \ref{fig:methodology_3}, respectively, on a diatomic lattice. For completeness, we show an example of a triatomic lattice in Fig.\ref{fig:methodology_4}. For the cases with the nonuniform breaking of TTS, multiple frequencies appear after the temporal interface. Hence, the temporal analogue of Snell's rule and Fernsel equations in literature are no longer applicable. We modify these rule theoretically and validated them numerically (Fig.\ref{fig:theory} and Fig.\ref{fig:Example_Fernsel equations}).    

\begin{figure*} 
\includegraphics
[scale= 1]{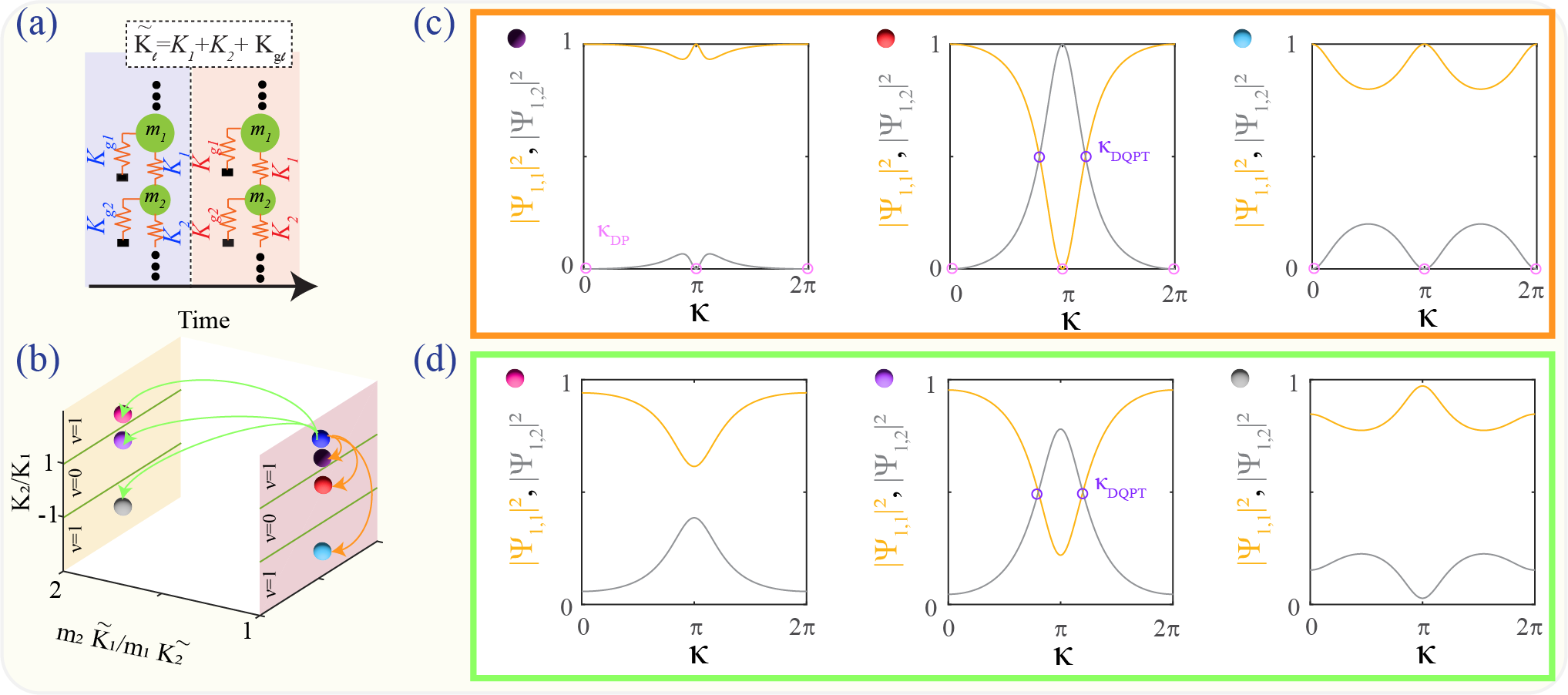} 
\caption{\label{fig:topology_SI}\textbf{Temporal interfaces as a platform for topology detection.} Diatomic structure with $m_1=0.5~m_2=0.1$ Kg and initially with $K_1=~0.5~K_2=~500$ N/m and $K_{g1}= 1000$ N/m and $K_{g2}=3500$ N/m. The structure is subjected to a temporal interface with nonuniform breaking of TTS. (a) Mathematical model. (b) Temporal interface changes system from $S_1$ (blue dot with $\frac{K_2}{K_1}=2$ and $\frac{m_2\Tilde{K_1}}{m_1 \Tilde{K_2}}=1$) to six different cases of $S_2$. Orange arrows represent $S_2$ satisfying chiral symmetry necessary condition. Green arrows represent $S_2$ not satisfying chiral symmetry necessary condition.(c) Analytical probability of each mode in $S_2$ after an incident wave on branch 1 of $S_1$ for the three cases with orange arrows in (b). (d) Analytical probability of each mode in $S_2$ after an incident wave on branch 1 of $S_1$ for the three cases with green arrows in (b).  }       
\end{figure*} 
\section{\large{Temporal interfaces as a platform for topology detection}}
Recent studies propose the utilization of the overlaps in Bloch mode shapes ($\Psi_{i,j}$) as a probe of topology in photonic lattices \cite{xu2025probing,wu2024edge}. These mode shape overlaps represent time refraction (i.e., same direction as the incident wave) and time reflection (i.e., opposite direction to the incident wave) coefficients ($|\Psi_{i,j}|^2$). Time reflection or refraction coefficients can also indicate the probability of each mode after a temporal interface. Time reflection or refraction coefficients vanish (i.e., zeros in Bloch mode shapes overlaps) at the wavenumbers of the band gap closing degenerate point(s) ($\kappa_{DP}$). The intersection between time reflection coefficient and time refraction coefficient ($|\Psi_{i,1}|^2=|\Psi_{i,2}|^2=0.5$~ at $\kappa_{DQPT}$) arise from distinct spatial topologies before and after a temporal interface (i.e., different winding numbers $\nu$). Such intersection ensures the existence of vanishing revival amplitude at critical times.

\begin{figure*} 
\includegraphics
[scale= 1]{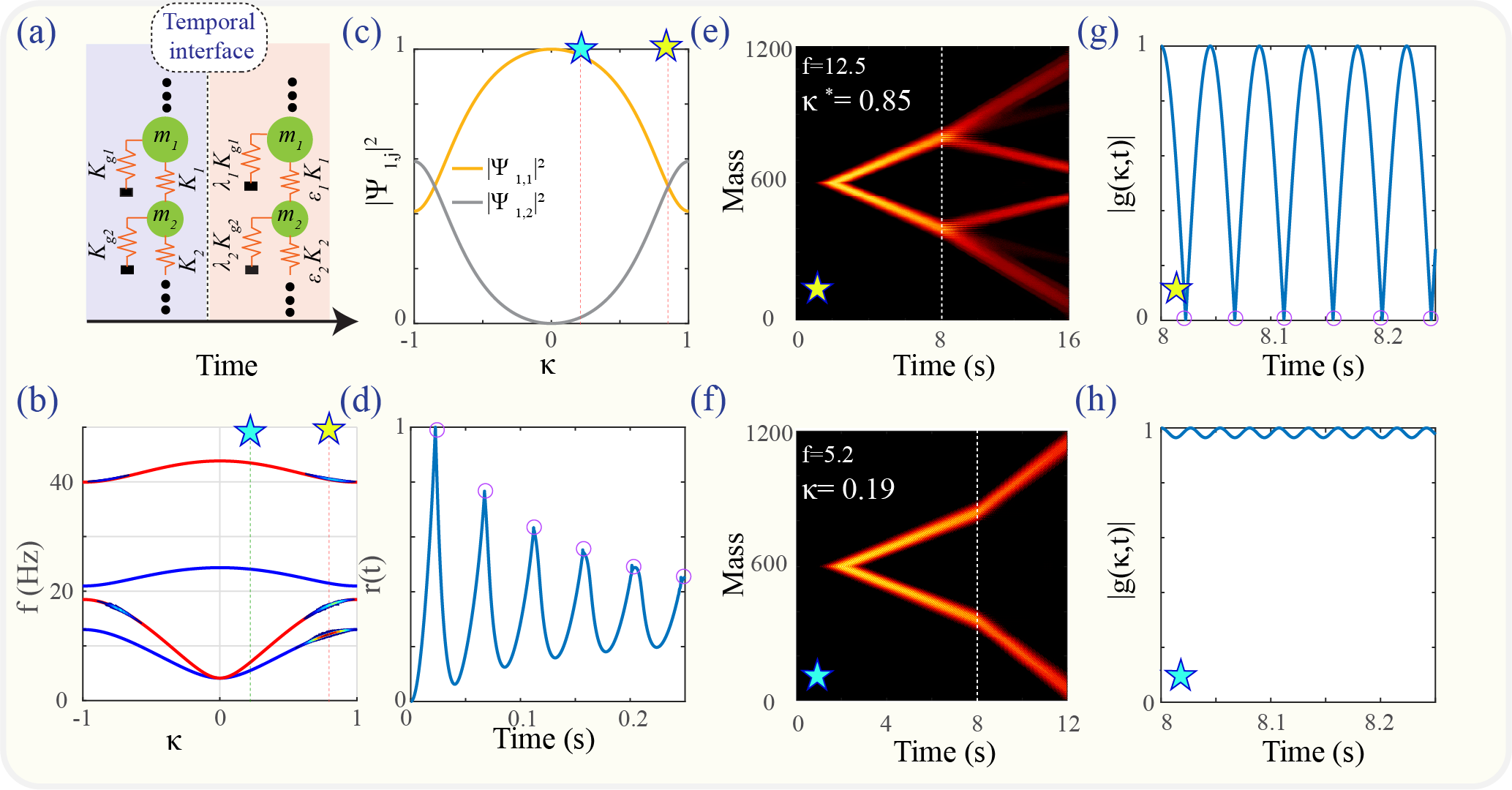} 
\caption{\label{fig:topology_SI2}\textbf{Temporal interfaces as a platform for topology detection.} Diatomic structure with $m_1=0.5~m_2=0.1$ Kg and initially with $K_1=~0.5~K_2=~500$ N/m and $K_{g1}=K_{g2}=100$ N/m. The structure is subjected to a temporal interface with nonuniform breaking of TTS ($\epsilon_1=8$ and $\epsilon_2=\lambda_1=\lambda_2=1$). (a) Mathematical model. (b) Analytical dispersion curves with 2D FFT overlay at $\kappa_{DQPT}=0.85\frac{\pi}{a}$. (c) Analytical probability of each mode in $S_2$ for an incident wave on mode 1 of $S_1$. (d) The rate function. Nonanalytic cusps represent dynamical quantum phase transition in the time boundary effect. (e) We excite a finite structure at the middle point with a wave packet $(\omega,\kappa_{DQPT})=(12.5 Hz, 0.85\frac{\pi}{a})$ which is corresponding to $(|\Psi_{1,1}|^2=|\Psi_{1,2}|^2=0.5)$. At $t=\tau=8~sec$, a sudden change from $S_1$ to $S_2$ takes place. Two frequencies propagate after the temporal interface. (f) The same as (e) but the excitation point is $(\omega,\kappa)=(5.2 Hz, 0.19\frac{\pi}{a})$ which is corresponding to a negligible contribution from mode 2 in $S2$ (i.e., $\Psi_{1,2}\approx0)$. A single frequency propagates after the temporal interface. (g) Revival " Loschmidt" amplitude for the initial state used in (e). The revival amplitude vanishes at the critical times (circles) which are the same times corresponding to cusps in (d). (h) The revival amplitude fro the initial state used in (f).    }
\end{figure*}

In phonic lattices, we need to modify such theory to accommodate unit-cell with different masses and/or with additional ground stiffness. We start with defining topological phase boundaries (i.e., band gap closes at point with degenerate frequencies). In a general configuration with a diatomic lattice (Fig.\ref{fig:topology_SI}-a), there are two topological phase boundaries (Fig.\ref{fig:topology_SI}-b):
(i)  when $\frac{K_2}{K_1}=1$ and $\frac{m_2 \Tilde{K_1}}{m_1 \Tilde{K_2}}=1$, $\kappa_{DP}$=$\pi$ and 
(ii) when $\frac{K_2}{K_1}=-1$ and $\frac{m_2 \Tilde{K_1}}{m_1 \Tilde{K_2}}=1$, $\kappa_{DP}$=$0,2\pi$. By tuning $S_1$ (before TI) and $S_2$ (after TI) such that $\frac{m_2 \Tilde{K_1}}{m_1 \Tilde{K_2}}=1$ for both states, Bloch mode shape overlaps vanish at three wavenumbers $\kappa_{DP}=0, \pi,$ and $2\pi$ (Fig.\ref{fig:topology_SI}-c). The topological invariant (i.e., the winding number $\nu$) can be calculated as:
\begin{equation}
    \nu=\int_{-\pi}^{\pi} B(\kappa) d\kappa 
\end{equation}
where $B(\kappa)$ is the Berry connection calculated as:
\begin{equation}
    B(\kappa)=\frac{1}{4\pi i}~tr[\mathbf{\sigma_3}~\mathbf{C}(\kappa)^{-1}~\partial_{\kappa}\mathbf {C}(\kappa)]
\end{equation}
where $\mathbf{\sigma_3}$ is the third pauli matrix and $\mathbf{C} (\kappa)$ is chiral matrix which we obtaine by eliminating the diagonal elements from the dynamic matrix. The winding number in the design plane  ($\frac{m_2 \Tilde{K_1}}{m_1 \Tilde{K_2}}=1$)   
equals 1 if $|K_2| > |K_1|$ and equals 0 if $|K_2| < |K_1|$ \cite{al2025inverse}. This divides the design plane to three regions: \newline
\textbf{Region 1.} topologically non-trivial with $\nu$=1 when $\frac{K_2}{K_1} >1$.
\newline
\textbf{Region 2.} topologically trivial with $\nu$=0 when $1>\frac{K_2}{K_1} >-1$.
\newline
\textbf{Region 3.} topologically non-trivial with $\nu$=1 when $\frac{K_2}{K_1} <-1$.
\newline
By selecting one state in region 1 or 3 and the other state in region 2, we get intersection points ($\kappa_{DQPT}$) between time refraction and time reflection coefficients, where the number of intersection points  within the region $[0,2\pi]$ equals 2 ($|\nu_1$-$\nu_2|)$ (Fig.\ref{fig:topology_SI}-c). The chiral symmetry condition is crucial for vanishing Bloch mode shapes overlaps (Fig.\ref{fig:topology_SI}-d). We emphasize that this theory is not applicable if we select states without satisfying the chiral symmetry necessary condition.    
 The revival amplitude ``Loschmidt amplitude" which quantifies the deviation of the time-evolved state from 
the initial condition is calculated as:
\begin{equation*}
    g(\kappa,t)=\sum_{j=1}^2~ |\Psi_{1,j} (\kappa)|^2~exp(-i\omega^{(S2)}_j(\kappa)~t)
\end{equation*}
where $\Psi_{1,j}$ is the overlap between modes (1 of $S_1$ and j of $S_2$) at a fixed wavenumber and $\omega^{(S2)}_j$ is the frequency on branch $j$ of $S2$ corresponding to the considered wavenumber. At the critical wavenumber $\kappa_{DQPT}$, revival amplitude vanishes periodically at critical times $t_{critical}$ where: $t_{critical}=(M-\frac{1}{2})~T_p$, $M=1,2,3,\dots$ and $T_p$ is the period (Fig.\ref{fig:topology_SI2}-g). By contrast, the revival amplitude
does not vanish for the initial state not at the critical wavenumber (Fig.\ref{fig:topology_SI2}-h). Moreover, we calculate the rate function $r(t)$ as follows:
\begin{equation*}
    r(t)=-(\frac{1}{N_\kappa})~ln ~(\prod_\kappa |g~(\kappa,t)|^2~)
\end{equation*}
where $N_\kappa$ is the number of wavenumbers considered in the BZ (i.e., wavenumber discretizations). At critical times, nonanalytic cusps or kinks appear in the rate function $r(t)$ (Fig.\ref{fig:topology_SI2}-d). The nonanalytic behavior at the critical times is the dynamical quantum phase transition
in the time boundary effect. The vanishing revival amplitude and the dynamical quantum phase
transition are ensured by the inequivalent band topology before and after the time boundary, and their appearances
are topologically-protected.

\end{document}